\documentclass[camera,letterpaper,nomarginnotes,nonarrowgutter]{jpaper}
\usepackage{mathptmx} 

\usepackage{fancyhdr}
\usepackage[normalem]{ulem}
\usepackage[sort]{cite}
\usepackage{microtype}
\usepackage{graphicx}
\usepackage{siunitx}
\usepackage{setspace}
\usepackage{flushend}
\usepackage{balance}

\usepackage{algorithm2e}


\usepackage{enumitem}
\usepackage{clipboard}

\usepackage[dvipsnames]{xcolor}
\usepackage[colorinlistoftodos,prependcaption,textsize=small]{todonotes}
\usepackage{xargs} 

\usepackage{tikz}
\usepackage[most]{tcolorbox}
\tcbuselibrary{breakable}
\tcbuselibrary{hooks}

\usepackage{multirow}

\usepackage{booktabs}

\usepackage{hhline}
\usepackage[bookmarks=true,breaklinks=true,hidelinks]{hyperref}

\usepackage{xcolor}
\definecolor{darkamber}{rgb}{0.6, 0.4, 0.0}
\definecolor{amber}{rgb}{1.0, 0.49, 0.0}
\definecolor{dkgreen}{rgb}{0.0,0.6,0.0}
\definecolor{gray}{rgb}{0.5,0.5,0.5}
\definecolor{dkgray}{rgb}{0.3,0.3,0.3}
\definecolor{mauve}{rgb}{0.58,0,0.82}
\definecolor{lightmauve}{rgb}{0.68,0.4,0.92}
\definecolor{chocolate}{rgb}{0.48, 0.25, 0.0}
\definecolor{amber}{rgb}{1.0,0.75,0}
\definecolor{dollarbill}{rgb}{0.52,0.73,0.4}
\definecolor{dkdkgreen}{rgb}{0,0.45,0}
\definecolor{gfored}{rgb}{0.580, 0.050, 0.211}
\definecolor{darkwarmgray}{rgb}{0.15, 0.050, 0.05}
\definecolor{ups-truck}{rgb}{0.53, 0.28, 0.21}
\definecolor{moegi}{rgb}{0.357, 0.537, 0.188}
\definecolor{darkcerulean}{rgb}{0.03, 0.27, 0.49}

\newif\ifrevision
\revisiontrue

\newif\ifcrone
\cronefalse

\newif\ifcrtwo
\crtwofalse

\newif\ifcrthree
\crthreefalse

\newif\ifcrfour
\crfourfalse

\newif\ifcrfive
\crfivefalse

\newif\ifcrsix
\crsixfalse

\newif\ifextone
\extonefalse

\newif\ifexttwo
\exttwofalse

\newif\ifextthree
\extthreefalse

\newif\ifonurready
\onurreadytrue

\ifonurready
    \definecolor{darkamber}{rgb}{0.2, 0.2, 1.0}
    \definecolor{amber}{rgb}{0.2, 0.2, 1.0}
    \definecolor{dkgreen}{rgb}{0.2, 0.2, 1.0}
    \definecolor{gray}{rgb}{0.2, 0.2, 1.0}
    \definecolor{dkgray}{rgb}{0.2, 0.2, 1.0}
    \definecolor{mauve}{rgb}{0.2, 0.2, 1.0}
    \definecolor{lightmauve}{rgb}{0.2, 0.2, 1.0}
    \definecolor{chocolate}{rgb}{0.2, 0.2, 1.0}
    \definecolor{amber}{rgb}{0.2, 0.2, 1.0}
    \definecolor{dollarbill}{rgb}{0.2, 0.2, 1.0}
    \definecolor{dkdkgreen}{rgb}{0.2, 0.2, 1.0}
    \definecolor{gfored}{rgb}{0.2, 0.2, 1.0}
    \definecolor{darkwarmgray}{rgb}{0.2, 0.2, 1.0}
    \definecolor{ups-truck}{rgb}{0.2, 0.2, 1.0}
    \definecolor{moegi}{rgb}{0.2, 0.2, 1.0}
    \definecolor{darkcerulean}{rgb}{0.2, 0.2, 1.0}
\fi

\newcommand{\agycrone}[1]{#1}
\newcommand{\agycronecomment}[1]{}
\newcommand{\hluocrone}[1]{#1}
\newcommand{\hluocronecomment}[1]{}
\newcommand{\omone}[1]{#1}
\newcommand{\omonecomment}[1]{}

\newcommand{\gfocomment}[1]{}
\newcommand{\mpo}[1]{{#1}}
\newcommand{\atbcrcomment}[1]{}
\newcommand{\mpcronecomment}[1]{}

\newcommand{\agycrtwo}[1]{#1}

\newcommand{\omtwo}[1]{#1}
\newcommand{\gfii}[1]{#1}

\newcommand{\agycrtwocomment}[1]{}
\newcommand{\hluocrtwocomment}[1]{}
\newcommand{\omtwocomment}[1]{}
\newcommand{\gftcomment}[1]{}
\newcommand{\atbcrtwocomment}[1]{}
\newcommand{\mpcrtwocomment}[1]{}
\newcommand\gfb[1][0]{}

\newcommand{\agycrthree}[1]{#1}
\newcommand{\agycrthreecomment}[1]{}
\newcommand{\omthree}[1]{#1}
\newcommand{\omthreecomment}[1]{}

\newcommand{\agycrfour}[1]{#1}
\newcommand{\agycrfourcomment}[1]{}
\newcommand{\omfour}[1]{#1}
\newcommand{\omfourcomment}[1]{}

\newcommand{\agycrfivecomment}[1]{}

\newcommand{\mpcrfivecomment}[1]{}

\newcommand{\omfivecomment}[1]{}

\newcommand{\agycrsixcomment}[1]{}

\newcommand{\mpcrsixcomment}[1]{}

\newcommand{\omsixcomment}[1]{}

\newcommand{\agyextone}[1]{#1}
\newcommand{\hluoextone}[1]{#1}
\newcommand{\agyextonecomment}[1]{}

\newcommand{\mpextonecomment}[1]{}

\newcommand{\omextonecomment}[1]{}

\newcommand{\agyexttwo}[1]{#1}

\newcommand{\agyexttwocomment}[1]{}

\newcommand{\mpexttwocomment}[1]{}
\newcommand{\omexttwo}[1]{#1}
\newcommand{\omexttwocomment}[1]{}

\newcommand{\agyextthree}[1]{#1}

\newcommand{\agyextthreecomment}[1]{}

\newcommand{\mpextthreecomment}[1]{}

\newcommand{\omextthreecomment}[1]{}

\ifextthree
  
  \renewcommand{\agyextthree}[1]{\textcolor{blue}{#1}}
  
  \renewcommand{\agyextthreecomment}[1]{\todo[linecolor=green,backgroundcolor=green!25,bordercolor=green, size=\tiny]{\textbf{@gy:} #1}}
  
  \renewcommand{\omextthreecomment}[1]{\todo[linecolor=red,backgroundcolor=red!25,bordercolor=red, size=\tiny]{\textbf{@om:} #1}}
\fi

\ifexttwo
  
  \renewcommand{\agyexttwo}[1]{\textcolor{blue}{#1}}
  
  \renewcommand{\agyexttwocomment}[1]{\todo[linecolor=green,backgroundcolor=green!25,bordercolor=green, size=\tiny]{\textbf{@gy:} #1}}
  \renewcommand{\omexttwo}[1]{\textcolor{gfored}{#1}}
  \renewcommand{\omexttwocomment}[1]{\todo[linecolor=red,backgroundcolor=red!25,bordercolor=red, size=\tiny]{\textbf{@om:} #1}}
\fi

\ifextone
  
  \renewcommand{\agyextone}[1]{\textcolor{blue}{#1}}
  \renewcommand{\hluoextone}[1]{\textcolor{blue}{#1}}
  \renewcommand{\agyextonecomment}[1]{\todo[linecolor=green,backgroundcolor=green!25,bordercolor=green, size=\tiny]{\textbf{@gy:} #1}}
  
  \renewcommand{\omextonecomment}[1]{\todo[linecolor=red,backgroundcolor=red!25,bordercolor=red, size=\tiny]{\textbf{@om:} #1}}
\fi

\ifcrsix

  \renewcommand{\agycrsixcomment}[1]{\todo[linecolor=green,backgroundcolor=green!25,bordercolor=green, size=\tiny]{\textbf{@gy:} #1}}
  
  \renewcommand{\omsixcomment}[1]{\todo[linecolor=red,backgroundcolor=red!25,bordercolor=red, size=\tiny]{\textbf{@om:} #1}}
\fi

\ifcrfive

  \renewcommand{\agycrfivecomment}[1]{\todo[linecolor=green,backgroundcolor=green!25,bordercolor=green, size=\tiny]{\textbf{@gy:} #1}}
  
  \renewcommand{\omfivecomment}[1]{\todo[linecolor=red,backgroundcolor=red!25,bordercolor=red, size=\tiny]{\textbf{@om:} #1}}
\fi

\ifcrfour
  
  \renewcommand{\agycrfour}[1]{\textcolor{blue}{#1}}
  \renewcommand{\agycrfourcomment}[1]{\todo[linecolor=green,backgroundcolor=green!25,bordercolor=green, size=\tiny]{\textbf{@gy:} #1}}
  \renewcommand{\omfour}[1]{\textcolor{gfored}{#1}}
  \renewcommand{\omfourcomment}[1]{\todo[linecolor=red,backgroundcolor=red!25,bordercolor=red, size=\tiny]{\textbf{@om:} #1}}
\fi

\ifcrthree
  
  \renewcommand{\agycrthree}[1]{\textcolor{blue}{#1}}
  \renewcommand{\agycrthreecomment}[1]{\todo[linecolor=green,backgroundcolor=green!25,bordercolor=green, size=\tiny]{\textbf{@gy:} #1}}
  \renewcommand{\omthree}[1]{\textcolor{gfored}{#1}}
  \renewcommand{\omthreecomment}[1]{\todo[linecolor=red,backgroundcolor=red!25,bordercolor=red, size=\tiny]{\textbf{@om:} #1}}
\fi

\ifcrone
\renewcommand{\agycrone}[1]{\textcolor{blue}{#1}}
\renewcommand{\agycronecomment}[1]{\todo[linecolor=green,backgroundcolor=green!25,bordercolor=green, size=\tiny]{\textbf{@gy:} #1}}

\renewcommand{\gfocomment}[1]{\todo[linecolor=green,backgroundcolor=green!25,bordercolor=green, size=\tiny]{\textbf{@gy:} #1}}
\renewcommand{\hluocrone}[1]{{\textcolor{moegi}{#1}}}
\renewcommand{\hluocronecomment}[1]{\todo[linecolor=blue,backgroundcolor=blue!25,bordercolor=blue, size=\tiny]{\textbf{@hluo:} #1}}
\renewcommand{\omone}[1]{\textcolor{blue}{#1}}
\renewcommand{\omonecomment}[1]{\todo[linecolor=red,backgroundcolor=red!25,bordercolor=red, size=\tiny]{\textbf{@om:} #1}}
\renewcommand{\atbcrcomment}[1]{\todo[linecolor=gray,backgroundcolor=gray!25,bordercolor=gray, size=\tiny]{\textbf{@atb:} #1}}
\renewcommand{\mpcronecomment}[1]{\todo[linecolor=amber,backgroundcolor=amber!25,bordercolor=amber, size=\tiny]{Minesh: #1}}
\renewcommand{\mpo}[1]{\textcolor{darkamber}{#1}}
\fi

\ifcrtwo
\renewcommand{\agycrtwo}[1]{\textcolor{blue}{#1}}

\renewcommand{\gftcomment}[1]{\todo[linecolor=green,backgroundcolor=green!25,bordercolor=green, size=\tiny]{\textbf{@gf:} #1}}

\renewcommand{\hluocrtwocomment}[1]{\todo[linecolor=blue,backgroundcolor=blue!25,bordercolor=blue, size=\tiny]{\textbf{@hluo:} #1}}
\renewcommand{\omtwo}[1]{\textcolor{gfored}{#1}}
\renewcommand{\gfii}[1]{\textcolor{gfored}{#1}}

\renewcommand{\omtwocomment}[1]{\todo[linecolor=red,backgroundcolor=red!25,bordercolor=red, size=\tiny]{\textbf{@om:} #1}}
\renewcommand{\agycrtwocomment}[1]{\todo[linecolor=green,backgroundcolor=green!25,bordercolor=green, size=\tiny]{\textbf{@gy:} #1}}

\renewcommand{\atbcrcomment}[1]{\todo[linecolor=gray,backgroundcolor=gray!25,bordercolor=gray, size=\tiny]{\textbf{@atb:} #1}}
\renewcommand{\mpcrtwocomment}[1]{\todo[linecolor=amber,backgroundcolor=amber!25,bordercolor=amber, size=\tiny]{Minesh: #1}}

\renewcommand{\gfb}[1]{[{\color{red}GF: #1}]}

\fi

\ifrevision

\fi

\newif\ifdraft
\draftfalse

\newcommand{\N}{\textcolor{red}{N}}

\newcommand{\td}[1]{}
\newcommand{\head}[1]{\noindent\textbf{#1.}}

\newcommand{\param}[1]{{#1}}

\newcommand{\agy}[1]{{#1}}
\newcommand{\agycomment}[1]{}
\newcommand{\js}[1]{{#1}}

\newcommand{\hluocomment}[1]{}

\ifdraft
\renewcommand{\param}[1]{\textcolor{red}{#1}}

\renewcommand{\agy}[1]{\textcolor{gfored}{#1}}
\renewcommand{\agycomment}[1]{\textcolor{gfored}{\textbf{[@gy:}#1\textbf{]}}}
\renewcommand{\js}[1]{\textcolor{darkamber}{#1}}

\renewcommand{\td}[1]{\textcolor{gfored}{\textbf{[TODO:}#1\textbf{]}}}

\renewcommand{\hluocomment}[1]{\textcolor{moegi}{\textbf{[@hluo: }#1\textbf{]}}}
\fi

\usepackage{tikz}
\usepackage{fancyhdr}
\usepackage{datetime} 

\settimeformat{ampmtime}

\newif\ifuseversions
\ifdraft
\useversionstrue
\fi
\newcommand{\versionnum}[0]{1.5}
\ifuseversions
  \def\parsepdfdatetime#1:#2#3#4#5#6#7#8#9{%
    \def\theyear{#2#3#4#5}%
    \def\themonth{#6#7}%
    \def\theday{#8#9}%
    \parsepdftime
  }

  \def\parsepdftime#1#2#3#4#5#6#7\endparsepdfdatetime{%
    \def\thehour{#1#2}%
    \def\theminute{#3#4}%
    \def\thesecond{#5#6}%
    \ifstrequal{#7}{Z}
    {%
      \def\thetimezonehour{+00}%
      \def\thetimezoneminute{00}%
    }%
    {%
      \parsepdftimezone#7%
    }%
  }

  \def\parsepdftimezone#1'#2'{%
    \def\thetimezonehour{#1}%
    \def\thetimezoneminute{#2}%
  }

  \newcommand*{\thetimezone}{\thetimezonehour:\thetimezoneminute}
  \expandafter\parsepdfdatetime\pdfcreationdate\endparsepdfdatetime

  \settimeformat{ampmtime}
  \newcommand{\version}[1]{\emph{Version #1 (Built:~\today~@ \currenttime~UTC\thetimezone)}}
  \AtBeginShipout
  {\AtBeginShipoutAddToBox{
      \begin{tikzpicture}[overlay, remember picture]
      \node[anchor=north] at (current page.north) {\textcolor{blue}{\vspace{3em}\version{\versionnum}}};    
      \end{tikzpicture}
  }}
\fi

\newcommand{\figref}[1]{Fig.~\ref{#1}}
\newcommand{\secref}[1]{§\ref{#1}}

\newcommand{\obsvref}[1]{Obsv.~\ref{#1}}
\newcommand{\obsvsref}[1]{Obsvs.~\ref{#1}}

\newcounter{obs}
\newcommand\observation[1]{%
   \refstepcounter{obs}
   \noindent
   \textbf{Observation \theobs.} \emph{#1}}

\newcommand{\squishlist}{
 \begin{list}{$\circ$}
  { \setlength{\itemsep}{0pt}
     \setlength{\parsep}{0pt}
     \setlength{\topsep}{0pt}
     \setlength{\partopsep}{0pt}
     \setlength{\leftmargin}{1em}
     \setlength{\labelwidth}{1em}
     \setlength{\labelsep}{0.5em} } }
\newcommand{\squishsublist}{
\begin{list}{$\rightarrow$}
 { \setlength{\itemsep}{0pt}
    \setlength{\parsep}{0pt}
    \setlength{\topsep}{-10em}
    \setlength{\partopsep}{-3pt}
    \setlength{\leftmargin}{1em}
    \setlength{\labelwidth}{1em}
    \setlength{\labelsep}{0.5em} } }

\newcommand{\squishend}{
  \end{list}  }

\makeatletter
\g@addto@macro{\normalsize}{%
  \setlength{\abovedisplayskip}{2pt plus 1pt minus 1pt}
  \setlength{\belowdisplayskip}{2pt plus 1pt minus 1pt}
  \setlength{\abovedisplayshortskip}{0pt}
  \setlength{\belowdisplayshortskip}{0pt}
  \setlength{\intextsep}{2pt plus 1pt minus 1pt}
  \setlength{\textfloatsep}{3pt plus 1pt minus 1pt}
  \setlength{\dbltextfloatsep}{3pt plus 1pt minus 1pt}
  \setlength{\skip\footins}{4pt plus 1pt minus 1pt}}
\makeatother

\newcommandx{\unsure}[2][1=]{\todo[linecolor=red,backgroundcolor=red!25,bordercolor=red,#1, size=\tiny]{#2}}
 \newcommandx{\change}[2][1=]{\todo[linecolor=blue,backgroundcolor=blue!25,bordercolor=blue,#1,size=\scriptsize]{#2}}
 \newcommandx{\feedback}[2][1=]{\todo[linecolor=yellow,backgroundcolor=yellow!25,bordercolor=yellow,#1]{#2}}
 \newcommandx{\improvement}[2][1=]{\todo[linecolor=Plum,backgroundcolor=Plum!25,bordercolor=Plum,#1]{#2}}
\newcommandx{\thiswillnotshow}[2][1=]{\todo[disable,#1]{#2}}
\newcommandx{\completedRevision}[2][1=]{\todo[disable,backgroundcolor=red,#1]{#2}}
\newcommandx{\dataSource}[2][1=]{\todo[disable,backgroundcolor=red,#1]{#2}}
 \newcommandx{\info}[2][1=]{\todo[linecolor=dollarbill,backgroundcolor=dollarbill!25,bordercolor=dollarbill,#1, size=\tiny]{#2}}

\definecolor{lightyellow}{rgb}{0.980, 0.956, 0.623}

\newcommand{\yboxbegin} {
	\begin{tcolorbox}[breakable, enhanced, frame hidden, colback=yellow!50]
}

\newcommand{\yboxend} {
	\end{tcolorbox}
}

\def\UrlBreaks{\do\/\do-\/\do.\/\do:}

\expandafter\def\expandafter\UrlBreaks\expandafter{\UrlBreaks
  \do\a\do\b\do\c\do\d\do\e\do\f\do\g\do\h\do\i\do\j
  \do\k\do\l\do\m\do\n\do\o\do\p\do\q\do\r\do\s\do\t
  \do\u\do\v\do\w\do\x\do\y\do\z\do\A\do\B\do\C\do\D
  \do\E\do\F\do\G\do\H\do\I\do\J\do\K\do\L\do\M\do\N
  \do\O\do\P\do\Q\do\R\do\S\do\T\do\U\do\V\do\W\do\X
  \do\Y\do\Z}
\usepackage{glossaries}

\newacronym{vdd}{$V_{DD}$}{supply voltage}
\newacronym{vpp}{$V_{PP}$}{wordline voltage}
\newacronym{vppmin}{$V_{PPmin}$}{the lowest \gls{vpp} at which the DRAM module can successfully communicate with the FPGA}
\newacronym{vwl}{$V_{PP}$}{wordline voltage}
\newacronym{gnd}{$GND$}{ground}
\newacronym{hcfirst}{$HC_{first}$}{the minimum aggressor row activation count \mpo{necessary} to cause a RowHammer bit flip}
\newacronym{rblast}{$r_{Blast}$}{blast radius}
\newacronym{ber}{$BER$}{the fraction of DRAM cells that experience a bit flip in a DRAM row}
\newacronym{nhc}{$N_{HC}$}{hammer count}
\newacronym{hc}{$HC$}{hammer count}
\newacronym{trcd}{$t_{RCD}$}{row activation latency}
\newacronym{tcl}{$t_{CL}$}{column access latency}
\newacronym{tcwl}{$t_{CWL}$}{column write latency}
\newacronym{trp}{$t_{RP}$}{precharge latency}
\newacronym{trcdmin}{$t_{RCDmin}$}{\agycrone{the minimum time delay \omtwo{required}}}
\newacronym{tras}{$t_{RAS}$}{charge restoration latency}
\newacronym{trasmin}{$t_{RASmin}$}{the minimum latency required}
\newacronym{trefw}{$t_{REFW}$}{refresh window}
\newacronym{vgs}{$V_{GS}$}{gate-to-source voltage}
\newacronym{vthresh}{$V_{TH}$}{the voltage threshold that the bitline voltage should exceed for the activation to be reliably completed}
\newacronym{kde}{KDE}{kernel density estimate}
\newacronym{ref}{refresh}{$REF$}
\newcommand{\berdecravg}{\SI{15.2}{\percent}}

\newcommand{\berdecrmax}{\SI{66.9}{\percent}}
\newcommand{\berincrmax}{\SI{11.7}{\percent}}

\newcommand{\hcfirstincravg}{\SI{7.4}{\percent}}
\newcommand{\hcfirstdecrmax}{\SI{9.1}{\percent}}
\newcommand{\hcfirstincrmax}{\SI{85.8}{\percent}}

\newcommand{\fracbersupportingrows}{\SI{81.2}{\percent}}
\newcommand{\fracberopposingrows}{\SI{15.4}{\percent}}
\newcommand{\fracbersmallchangerowsmfrA}{\SI{49.6}{\percent}}

\newcommand{\frachcfirstsupportingrows}{\SI{69.3}{\percent}}
\newcommand{\frachcfirstopposingrows}{\SI{14.2}{\percent}}

\newcommand{\frachcfirstsupportingrowsmfrA}{\SI{50.9}{\percent}}

\newcommand{\frachcfirstsupportingrowsmfrC}{\SI{83.5}{\percent}}
\newcommand{\bersmallchange}{\SI{2}{\percent}}

\newcommand{\minnormbermfrA}{{0.43}}
\newcommand{\maxnormbermfrA}{{1.11}}
\newcommand{\minnormbermfrB}{{0.33}}
\newcommand{\maxnormbermfrB}{{1.03}}
\newcommand{\minnormbermfrC}{{0.74}}
\newcommand{\maxnormbermfrC}{{0.94}}
\newcommand{\minnormhcfirstmfrA}{{0.94}}
\newcommand{\maxnormhcfirstmfrA}{{1.52}}
\newcommand{\minnormhcfirstmfrB}{{0.92}}
\newcommand{\maxnormhcfirstmfrB}{{1.86}}
\newcommand{\minnormhcfirstmfrC}{{0.91}}
\newcommand{\maxnormhcfirstmfrC}{{1.35}}

\newcommand{\numchips}{272}
\newcommand{\numreliablechips}{208}
\newcommand{\numunreliablechips}{64}
\newcommand{\numunreliablemodules}{5}
\newcommand{\numreliablemodules}{25}
\newcommand{\nummodules}{30}
\newcommand{\trcdguardbandreduction}{\SI{21.9}{\percent}}

\def\BibTeX{{\rm B\kern-.05em{\sc i\kern-.025em b}\kern-.08em
    T\kern-.1667em\lower.7ex\hbox{E}\kern-.125emX}}

\fancypagestyle{firstpage}{
  \fancyhf{}

  \fancyfoot[C]{\thepage}
}

\newcommand{\affilETH}{$^1$}
\newcommand{\affilCESGA}{$^2$}
\newcommand{\affilLois}{$^{1,2}$}

\title{Understanding \agycrone{RowHammer} Under Reduced Wordline Voltage:\\An Experimental Study Using Real DRAM Devices}
\author{\vspace{-18pt}\\%
\fontsize{11}{12}\selectfont%
{A. Giray Ya\u{g}l{\i}k\c{c}{\i}\affilETH{}}\quad%
{Haocong Luo\affilETH{}}\quad%
{Geraldo F. de Oliviera\affilETH{}}\quad%
{Ataberk Olgun\affilETH{}}\quad%
{Minesh Patel\affilETH{}}\quad%
\vspace{-1pt}\\%
\fontsize{11}{12}\selectfont%
{Jisung Park\affilETH{}}\quad%
{Hasan Hassan\affilETH{}}\quad
{Jeremie S. Kim\affilETH{}}\quad%
{Lois Orosa\affilLois{}}\quad%
{Onur Mutlu\affilETH{}}%
\vspace{0pt}\\%
{\fontsize{10}{11}\selectfont
\qquad\qquad
\affilETH\emph{ETH Z{\"u}rich}
\qquad\qquad
\affilCESGA\emph{Galicia Supercomputing Center (CESGA)}%
}
\vspace{-12pt}}

\glsdisablehyper
\begin{document}
\bstctlcite{IEEEexample:BSTcontrol}

\maketitle
\thispagestyle{firstpage}
\pagestyle{firstpage}

\setstretch{0.9225}
\begin{abstract}
RowHammer is a circuit-level {DRAM vulnerability}, where {repeatedly} {activating and precharging} a DRAM row, and thus alternating {the voltage of a row's wordline between low and high voltage levels}, can cause bit flips in physically nearby rows.
{Recent {DRAM} chips are more vulnerable to RowHammer: {{with technology node scaling,} the minimum number of activate-precharge cycles to induce a RowHammer bit flip reduces {and} the RowHammer bit error rate increases.}} {Therefore, it is critical to develop {effective and scalable} approaches {to} protect modern DRAM {systems} against RowHammer.}
{To {enable such solutions,}
it is essential to develop a deeper understanding of the RowHammer vulnerability of modern DRAM chips. However,}
{even though the voltage {toggling} on a wordline is a key determinant of RowHammer vulnerability, \emph{no} prior work experimentally demonstrates the effect of \gls{vpp} on the RowHammer vulnerability.} {Our work closes this gap in understanding.}

{This is} the first work to experimentally demonstrate 
on {\numchips{}} real DRAM chips that {lowering} 
\gls{vpp} reduces a DRAM chip's RowHammer vulnerability{. We show that
{lowering}
\gls{vpp} 1)~increases} the {number of {activate-precharge cycles} 
{needed} to induce} a {RowHammer} bit flip {by up to \hcfirstincrmax{} with an average of \hcfirstincravg{} across all tested chips} and 2)~{decreases the} RowHammer bit error rate {by up to \berdecrmax{} with an average of \berdecravg{} across all tested chips}.
{At the same time,} reducing \gls{vpp} marginally {worsens} a DRAM cell's access latency, charge restoration, and {data} retention {time} 
within the guardbands {of system-level nominal timing parameters} for {\numreliablechips{} out of \numchips{} tested chips.} 
{We conclude that} {reducing \gls{vpp} is a {promising} strategy for reducing a DRAM chip's RowHammer vulnerability without requiring modifications to DRAM chips.}

\end{abstract}
\glsresetall
\section{Introduction}
\label{sec:introduction}

{Manufacturing process technology scaling continuously increases}
DRAM storage density by reducing \mpo{circuit component sizes and enabling tighter packing of DRAM cells.}
{S}uch {advancements} 
{reduce} {DRAM chip cost but worsen} DRAM reliability~\cite{mutlu2013memory, meza2015revisiting}. Kim et al.~\cite{kim2014flipping} show that modern DRAM chips are susceptible to {a read disturbance effect, called \emph{RowHammer}}, where {repeatedly} {activating and precharging}
a DRAM row (i.e., \emph{aggressor row}) {many times} (i.e., \emph{hammering} {the {aggressor} row}) can cause bit flips in {physically nearby} rows (i.e., {\emph{victim rows}}) {at consistently predictable bit locations}~\cite{redeker2002investigation, kim2014flipping, aichinger2015ddr, park2016experiments, park2016statistical, mutlu2017rowhammer, mutlu2019rowhammer, yang2019trap, kim2020revisiting, orosa2021deeper, qureshi2021rethinking, saroiu2022price, walker2021ondramrowhammer}.

{{M}any works~{\cite{seaborn2015exploiting, van2016drammer, gruss2016rowhammer, razavi2016flip, pessl2016drama, xiao2016one, bosman2016dedup, bhattacharya2016curious, qiao2016new, jang2017sgx, aga2017good,
mutlu2017rowhammer, tatar2018defeating ,gruss2018another, lipp2018nethammer,
van2018guardion, frigo2018grand, cojocar2019eccploit,  ji2019pinpoint, mutlu2019rowhammer, hong2019terminal, kwong2020rambleed, frigo2020trrespass, cojocar2020rowhammer, weissman2020jackhammer, zhang2020pthammer, rowhammergithub, yao2020deephammer, deridder2021smash, hassan2021utrr, jattke2022blacksmith, marazzi2022protrr, tol2022toward,burleson2016invited, brasser2017can,qureshi2021rethinking, orosa2021deeper, redeker2002investigation, saroiu2022price, yang2019trap, walker2021ondramrowhammer, park2016statistical, kim2020revisiting, kim2014flipping, park2016experiments} demonstrate that RowHammer is a serious security vulnerability that can be exploited to mount system{-}level attacks, such as escalating} privilege or {leaking} private data.}
To make matters worse, recent experimental studies {on real DRAM chips}~\cite{mutlu2017rowhammer, mutlu2019rowhammer, frigo2020trrespass, cojocar2020rowhammer, kim2020revisiting, kim2014flipping, hassan2021utrr, orosa2021deeper} {find} that the RowHammer vulnerability is 
more severe in newer DRAM chip generations.
For example{,} 
{1)~\gls{hcfirst}} is {\emph{only} {4.8K and 10K}  
for {some} newer {LPDDR4 and DDR4 DRAM} chips (manufactured in {201{9--2020}}), which is
{$14.4\times$ and $6.9\times$} 
lower than the \gls{hcfirst} of {69.2}K for {some} older} DRAM chips (manufactured in {2010--2013}{)~\cite{kim2020revisiting}; and 
2)~\gls{ber} after hammering two aggressor rows for $30K$ times is $2\times10^{-6}$ for {some} newer DRAM chips {from 2019--2020}, which is $500\times$ larger than that for {some other} older chips {manufactured in 2016--2017} ($4\times10^{-9}$)~\cite{kim2020revisiting}.}
As the RowHammer vulnerability worsens, ensuring RowHammer-safe operation {becomes} more expensive {across a broad range of system-level design metrics, including} performance overhead, energy consumption, and hardware complexity~{\cite{mutlu2014research, mutlu2017rowhammer, mutlu2019rowhammer, kim2020revisiting, yaglikci2021blockhammer, park2020graphene, orosa2021deeper, hassan2021utrr,yaglikci2021security,frigo2020trrespass}}.

To find effective and efficient solutions for RowHammer, it is {essential} to develop a deeper understanding of the RowHammer vulnerability of modern DRAM chips~{\cite{orosa2021deeper, mutlu2017rowhammer, mutlu2019rowhammer}}. Prior {works~\cite{kim2014flipping, redeker2002investigation, mutlu2017rowhammer, mutlu2019rowhammer, walker2021ondramrowhammer, yang2019trap, park2016statistical, park2016experiments, orosa2021deeper, kim2020revisiting} hypothesize that the RowHammer vulnerability originates from} circuit-level interference {between 1)~wordlines that are physically nearby each other and 2)~between a wordline and physically nearby DRAM cells.} 
Existing {circuit-level} models~{\cite{park2016statistical,yang2019trap,walker2021ondramrowhammer}}  
suggest that {toggling of the} voltage on a wordline is a {key} 
{determinant of how much repeated aggressor row activations disturb physically nearby circuit components.}
However, it is still {unclear} 1)~how {the magnitude of {the} \gls{vpp} affects} {modern DRAM chips'} RowHammer vulnerability and 2)~whether it is possible to reduce 
RowHammer vulnerability {by} {reducing} \gls{vpp}{, without {significantly} worsening other issues related to reliable DRAM operation}. 
{{Therefore,} \textbf{our goal} 
is to {experimentally} understand {how \gls{vpp} affects} RowHammer vulnerability and DRAM {operation}.}

\noindent
{{\textbf{Our Hypothesis.} 
{W}e hypothesize that {lowering}  
\gls{vpp} can reduce {RowHammer vulnerability} 
without 
{significantly impacting
{reliable} DRAM {operation}.}
To test this hypothesis,}
we experimentally demonstrate how 
RowHammer vulnerability varies with \gls{vpp} {by conducting} rigorous {experiments} on \numchips{} real DDR4 DRAM chips {from three major {DRAM} vendors}.} {To} isolate the effect o{f} \gls{vpp} and to avoid failures in DRAM chip I/O circuitry, we scale {\emph{only}} \gls{vpp} {and} supply the rest of the {DRAM} circuitry {using} {the} nominal \gls{vdd}. 

\noindent
{\textbf{Key Findings.}} 
{O}ur experimental results {yield} six novel observations {about} {\gls{vpp}'s effect on RowHammer} (\secref{sec:vpp_with_rh}). {Our key observation is} that a DRAM chip's RowHammer vulnerability {reduces} by scaling down \gls{vpp}{:} 1)~\gls{hcfirst} increases by \hcfirstincravg{} {(\hcfirstincrmax{})}\js{,} and 2)~the {\gls{ber}} 
caused by a RowHammer attack reduces by \berdecravg{} {(\berdecrmax{})}\omone{,} on average {({at} max)} {across all tested DDR4 DRAM chips}.

{To investigate the potential {adverse} effects of reducing \gls{vpp} on {reliable} DRAM {operation},}
we conduct experiments {using {both} real DDR4 DRAM chips} and SPICE~\cite{nagel1973spice} simulations {that} measure how reducing \gls{vpp} affects a DRAM {cells'} 1)~{row activation} latency, 2)~charge restoration process, and 3)~{data} retention time. {Our measurements yield} {nine} novel observations {(\secref{sec:sideeffects})}. {{We make two} key observation{s:}} {First,} \gls{vpp} reduction {only \emph{marginally} worsens} {the} access latency, charge restoration process, and {data} retention time of {most} DRAM chip{s}{:} 
{\numreliablechips{} out of \numchips{}} tested DRAM {chips}
reliably operate using nominal timing parameters due to {the built-in \emph{safety {margins}} {(i.e., \emph{guardbands})} in nominal timing parameters that DRAM manufacturers already provide. {Second, \numunreliablechips{} DRAM chips that exhibit erroneous behavior at reduced \gls{vpp} can reliably operate using 1)~a {longer} row activation latency, i.e., \SI{24}{\nano\second} / \SI{15}{\nano\second} for 48 / 16 chips (\secref{sec:sideeffects_trcd}), 2)~simple single-error-correcting codes~\cite{hamming1950error} (\secref{sec:sideeffects_retention}), {or} 3)~{doubling the} refresh {rate \emph{only} for} \SI{16.4}{\percent} of DRAM rows.}} 

We make the following \omone{major} contributions in this paper. 

\squishlist
    \item We present the first experimental {RowHammer characterization study} under reduced wordline voltage (\gls{vpp}). 
    \item Our experiments on \numchips{} real DDR4 DRAM chips show that {when a DRAM module is operated at a {reduced} \gls{vpp},}
    {an attacker 1)~needs to hammer a row {in the module more {times} (by \hcfirstincravg{}  /} {\hcfirstincrmax{})} to induce a bit flip, and 2)~can cause fewer (\berdecravg{} / {\berdecrmax{})} {RowHammer} bit flips {in the module} {(on average /} {at max{imum}} across {all} tested {modules)}.}
    \item We {present the first {experimental} study of how} reducing \gls{vpp} {affects} DRAM access latency, charge restoration process, and {data} retention time{.}
    \item {Our experiments} {on real DRAM chips} show that reducing \gls{vpp} {slightly} {worsens} DRAM access latency, {charge restoration process}, and {data} retention time{.} 
    {Most (}\numreliablechips{} out of \numchips{}{)} DRAM chips reliably {operate} under reduced \gls{vpp}{, while the remaining \numunreliablechips{} chips reliably operate using increased row activation latency, simple error correcting codes, {or doubling the refresh rate \emph{only} for \SI{16.4}{\percent} of the rows.}}
\squishend

\glsresetall
\section{Background}
\label{sec:background}

{We provide} a high-level overview of DRAM design and operation as relevant to our work. For a more detailed overview, we refer the interested reader to prior works{~\cite{keeth2001dram, keeth2007dram, lee2013tiered, lee2015adaptive,seshadri2013rowclone, chang2017understanding,chang2016low, vampire2018ghose,hassan2016chargecache,hassan2017softmc,hassan2019crow, khan2014efficacy, khan2016parbor, khan2017detecting,kim2018solar,kim2018dram, kim2014flipping,kim2016ramulator,lee2017design,lee2015decoupled,liu2013experimental,liu2012raidr, luo2020clrdram, patel2017reaper, qureshi2015avatar, seshadri2017ambit}}.

\subsection{DRAM Background}
\begin{figure}[h]
    \centering
    \includegraphics[width=\linewidth]{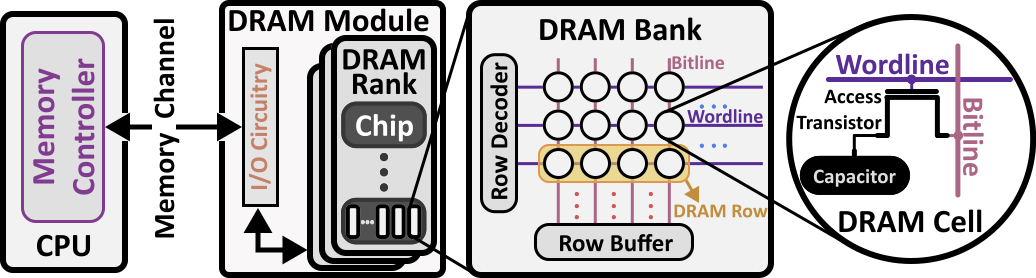}
    \caption{Organization of a typical modern DRAM module.}
    \label{fig:dram_background}
\end{figure}

\noindent\textbf{{DRAM Organization.}}
\figref{fig:dram_background} illustrates a DRAM module's hierarchical organization.
At the lowest level of the hierarchy, a single DRAM cell comprises {1)~}a \emph{storage capacitor} that stores a single bit of data encoded using the charge level {in} the capacitor and 2)~an \emph{access transistor} that is used to read from and write to the storage capacitor. The DRAM cell is connected to a \emph{bitline} that is used to access the data stored in the cell and a \emph{wordline} that controls access to the cell. 

DRAM cells are {organized as} a two-dimensional array to form a \emph{bank}. Each cell in a bank is addressed by its \emph{row} and \emph{column}. 
{Each DRAM cell {in} a DRAM row is connected to a {common \emph{wordline}} via {its} \emph{access transistor}. {A \emph{bitline}} connects a column of DRAM cells to a DRAM {\emph{sense amplifier}} to read {or} write data.} {A row of sense amplifiers is called a \emph{row buffer.}} 
Multiple (e.g., 16~\cite{jedec2017ddr4}) DRAM banks are  
{put together} to form a single DRAM \emph{chip}.
Multiple chips form a \emph{rank}{.} {Chips in a rank} operate in lock-step such that {each chip serves a portion of the data {for} each DRAM access.}
A DRAM module may have one or more ranks, communicating with the memory controller over the {\emph{memory channel}}.

\noindent\textbf{DRAM Operation.}
The memory controller services main memory requests {using} three key operations.

\emph{1)~Row Activation.} The memory controller {sends an $ACT$ command along with} a row address {to} a bank, and the DRAM chip asserts the corresponding wordline {to activate the DRAM row}. 
{Asserting a wordline {connects} each cell capacitor in the activated row to its corresponding bitline, perturbing the bitline voltage. Then, the sense amplifier senses and amplifies the voltage perturbation until the cell charge is restored. The data is accessible when the bitline voltage is amplified to a certain {level}. {The latency {from} the start
of row activation {until {the data is {reliably readable}}} is called \emph{\gls{trcd}}.}
A DRAM cell loses its charge during row activation, and thus its initial charge needs to be restored before the row is closed. {The latency {from} the start
of row activation {until {the completion of}} the DRAM cell's charge {restoration} is called \emph{\gls{tras}}.} 
DRAM manufacturers provide a built-in safety margin in the nominal timing parameters to account for the worst-case latency in {\gls{trcd} and \gls{tras}} operations~\cite{lee2015adaptive, chang2016understanding, chang2017understanding}.}

\emph{2)~Read/Write.} The memory controller {sends a $RD$/$WR$ command along with} a column {address} to perform a read or write {to the activated row in the DRAM bank. A $RD$ command {serves} data from the row buffer {to the memory channel. A} $WR$ command writes data into the row buffer, which subsequently {modifies} the {data stored in the} DRAM cell.} 
{The latency of performing a read/write operation is called \emph{\gls{tcl} / \gls{tcwl}}.}

\emph{3)~Precharge.} The memory controller {sends a $PRE$ command to an active bank. {The DRAM chip de-asserts the active row's} wordline and {precharges} the bitlines to} prepare the DRAM bank for a new row activation. \gfii{ The timing parameter for precharge is called {\emph{\gls{trp}}}, which is the latency between issuing a $PRE$ command and when the DRAM bank is ready for a new row activation.}

\noindent \textbf{DRAM Refresh.} A DRAM cell {inherently leaks charge and thus can retain data for \hluocrone{only} a limited amount of time, called \emph{\agycrone{data} retention time}.}
To prevent data loss due to \hluocrone{such} leakage, the memory controller \gfii{periodically} issues \agycrone{\emph{\gls{ref}} commands that ensure every DRAM cell is refreshed at a fixed interval, called \emph{\gls{trefw}}} \hluocrone{(e.g., every \SI{64}{\milli\second}~\gfii{\cite{jedec2012ddr3, jedec2017ddr4,jedec2020ddr5}} or \SI{32}{\milli\second}~\cite{jedec2015lpddr4}).}

\subsection{DRAM Voltage Control}
\label{sec:background:dramvoltage}
Modern DRAM chips {(e.g., DDR4~\cite{jedec2017ddr4}, DDR5~\cite{jedec2020ddr5}, GDDR5X~\cite{jedec2016gddr5x}, and GDDR6~\cite{jedec2021gddr6} {standard compliant ones})} use two separate voltage rails: {1)~\gls{vdd}, which is used to operate the core DRAM array and peripheral circuitry (e.g., the sense amplifiers, row/column decoders, precharge and I/O logic), and 2)~\gls{vpp}, which is exclusively used to assert {a} wordline during a DRAM row activation.} 
\gls{vpp} is generally significantly higher (e.g., 2.5V~\cite{salp, micron2016trr,DS_MTA18A,micron2014ddr4}) than \gls{vdd} (e.g., 1.25--1.5V~\cite{salp, micron2016trr,DS_MTA18A,micron2014ddr4}) in order to ensure 1)~full activation of {all} access {transistors of a row} when the wordline is {asserted} and 2)~low leakage when the wordline is {de-asserted}. 
\gls{vpp} is internally generated from \gls{vdd} in older DRAM chips (e.g., DDR3~\cite{jedec2012ddr3}). However, newer DRAM chips (e.g., DDR4 onwards~{\cite{jedec2017ddr4, jedec2020ddr5, jedec2016gddr5x, jedec2021gddr6}}) expose \emph{both} \gls{vdd} and \gls{vpp} rails to external pins, allowing \gls{vdd} and \gls{vpp} to be independently driven with different voltage {sources}. 

\subsection{The RowHammer Vulnerability}
\label{sec:background_rowhammer}
Modern DRAM is susceptible to a circuit-level vulnerability known as RowHammer~{\cite{redeker2002investigation, kim2014flipping, aichinger2015ddr, park2016experiments, park2016statistical, mutlu2017rowhammer, mutlu2019rowhammer, yang2019trap, kim2020revisiting, orosa2021deeper, qureshi2021rethinking, saroiu2022price, walker2021ondramrowhammer}}, where a cell's stored data can be corrupted by {repeatedly activating} physically nearby {(aggressor)} rows{. RowHammer} {results} in unwanted software-visible bit flips {and breaks memory isolation~\cite{kim2014flipping, mutlu2019rowhammer,mutlu2017rowhammer}}. RowHammer {poses} a significant {threat to system security{,} reliability{, and DRAM technology scaling.}} First, RowHammer lead{s} to data corruption{,} system crashes{, and security attacks} if not appropriately mitigated. {Many} prior works~\cite{seaborn2015exploiting, van2016drammer, gruss2016rowhammer, razavi2016flip, pessl2016drama, xiao2016one, bosman2016dedup, bhattacharya2016curious, qiao2016new, jang2017sgx, aga2017good,
mutlu2017rowhammer, tatar2018defeating ,gruss2018another, lipp2018nethammer,
van2018guardion, frigo2018grand, cojocar2019eccploit,  ji2019pinpoint, mutlu2019rowhammer, hong2019terminal, kwong2020rambleed, frigo2020trrespass, cojocar2020rowhammer, weissman2020jackhammer, zhang2020pthammer, rowhammergithub, yao2020deephammer, deridder2021smash, hassan2021utrr, jattke2022blacksmith, marazzi2022protrr, tol2022toward, burleson2016invited, brasser2017can} show that RowHammer can be exploited to mount system-level attacks to compromise system security (e.g., to acquire root privileges {or leak private data}).
Second, RowHammer {vulnerability worsens} {as DRAM technology scales to smaller node sizes}~\cite{mutlu2017rowhammer, mutlu2019rowhammer, frigo2020trrespass, cojocar2020rowhammer, kim2020revisiting, kim2014flipping, hassan2021utrr, orosa2021deeper}. This is because process technology shrinkage reduces the size of circuit elements, exacerbating charge leakage paths in and around each DRAM cell. Prior work{s}~\cite{kim2020revisiting, frigo2020trrespass, hassan2021utrr, orosa2021deeper} experimentally demonstrate {with modern DRAM chips that} RowHammer is and will continue to be an increasingly significant reliability, security, and safety problem going forward{~\cite{mutlu2017rowhammer,mutlu2019rowhammer}{, given that \gls{hcfirst} is \emph{only} {4.8K} in modern DRAM chips~\cite{kim2020revisiting} and {it} continues {to reduce}.}}

{We describe two major error mechanisms that lead to Row{H}ammer, as explained by prior works~\cite{redeker2002investigation, park2016statistical, yang2019trap, walker2021ondramrowhammer, ryu2017overcoming, park2016experiments, sakurai1993closed}: 1) \emph{electron injection / diffusion / drift} and 2) \emph{capacitive crosstalk}.}
{The} \emph{electron injection / diffusion / drift}
{mechanism} create{s} temporary charge leakage paths that degrade the voltage of a cell's storage capacitor~\cite{walker2021ondramrowhammer, yang2019trap, ryu2017overcoming, park2016experiments}. A larger voltage difference between {a wordline and a DRAM cell or between two wordlines}
{exacerbates {the electron injection / diffusion / drift}} error mechanism.
{The} \emph{capacitive crosstalk} {mechanism} exacerbates charge leakage paths in and around a DRAM cell's capacitor~\cite{walker2021ondramrowhammer, ryu2017overcoming, redeker2002investigation, sakurai1993closed} {due to the} {parasitic} capacitance between {two wordlines or between a wordline and a DRAM cell}.

\subsection{Wordline Voltage\agy{'s Impact} on DRAM Reliability}

\head{RowHammer}
As explained in \secref{sec:background_rowhammer}{,}
a larger \gls{vpp} exacerbates both {electron injection / diffusion / drift}
and capacitive crosstalk mechanisms. Therefore, we hypothesize that the RowHammer vulnerability of a DRAM chip increases {as} \gls{vpp} {increases}. Unfortunately, {there is no prior work that tests this hypothesis and quantifies the effect of \gls{vpp} on real DRAM chips’ RowHammer vulnerability.}
\secref{sec:motivation} discusses this hypothesis in further detail, and \secref{sec:vpp_with_rh} experimentally examines the effects of changing \gls{vpp} on the RowHammer vulnerability of real DRAM chips. 

\head{{Row Activation and Charge Restoration}}
{An access transistor turns on (off) {when its gate voltage is higher (lower) than a threshold.}} {An access transistor's gate is connected to a wordline (\figref{fig:dram_background}) and driven by \gls{vpp} (ground) when the row is {activated} ({precharged}).\footnote{{To increase DRAM cell retention time, modern DRAM {chips} may apply a negative voltage to the wordline~\cite{Frank2001Devicescaling, Lee2011SimultaneousReverse} when {the wordline} is not {asserted}{. Doing so} {reduces} the leakage current {and this improves data retention}.}}} 
{Between \gls{vpp} and ground, a larger access transistor gate voltage} forms a stronger channel between the bitline and the capacitor.
{A} strong channel allows {fast DRAM row activation and full charge restoration.}
Based on these properties, we hypothesize that a \emph{larger} \gls{vpp} provides \emph{smaller} row activation latency and increased {data} retention time, leading to {more reliable DRAM operation}.\footnote{{Increasing/decreasing \gls{vpp} does \emph{not} affect the reliability of $RD$/$WR$ and $PRE$ operations since the DRAM {circuit components} involved in these operations are powered using \emph{only} \gls{vdd}.}}
Unfortunately, {there is \emph{no} prior work that tests this hypothesis and quantifies \gls{vpp}'s effect on real DRAM chips' {reliable operation (i.e.,} row activation and charge restoration {characteristics)}.}
{\secref{sec:sideeffects} {studies} {the effect of} reduced \gls{vpp} {on DRAM operation reliability} using both {real-device characterizations} and SPICE~\cite{nagel1973spice, ltspice} simulations.}

\section{Motivation}
\label{sec:motivation}

RowHammer is a critical {vulnerability} for modern DRAM-based computing platforms~{\cite{redeker2002investigation, kim2014flipping, park2016experiments, park2016statistical, mutlu2017rowhammer, mutlu2019rowhammer, yang2019trap, kim2020revisiting, orosa2021deeper, qureshi2021rethinking, saroiu2022price, seaborn2015exploiting, aichinger2015ddr, 
van2016drammer, gruss2016rowhammer, razavi2016flip, xiao2016one, bosman2016dedup, bhattacharya2016curious, qiao2016new, jang2017sgx, aga2017good, tatar2018defeating, lipp2018nethammer,
van2018guardion, frigo2018grand, cojocar2019eccploit,  ji2019pinpoint, hong2019terminal, kwong2020rambleed, frigo2020trrespass, cojocar2020rowhammer, weissman2020jackhammer, zhang2020pthammer, rowhammergithub, yao2020deephammer, deridder2021smash, hassan2021utrr, jattke2022blacksmith, marazzi2022protrr, tol2022toward,
burleson2016invited, gruss2018another, walker2021ondramrowhammer, brasser2017can, pessl2016drama}.} 
{Many prior works~\cite{aichinger2015ddr, AppleRefInc, aweke2016anvil, kim2014flipping, kim2014architectural,son2017making, lee2019twice, you2019mrloc, seyedzadeh2018cbt, van2018guardion, konoth2018zebram, park2020graphene, yaglikci2021blockhammer, kang2020cattwo, bains2015row, bains2016distributed, bains2016row, brasser2017can, gomez2016dummy, jedec2017ddr4,hassan2019crow, devaux2021method, ryu2017overcoming, yang2016suppression, yang2017scanning, gautam2019row, yaglikci2021security, qureshi2021rethinking, greenfield2012throttling,
marazzi2022protrr, saileshwar2022randomized} propose RowHammer mitigation {mechanisms} that {aim to} prevent RowHammer bit flips. Unfortunately, RowHammer solutions need to consider a {large} {number of} design space {constraints} that include cost, performance impact, energy and power overheads, hardware complexity, {technology} scalability, security guarantees, and changes to existing DRAM standards and interfaces. Recent works~{\cite{kim2020revisiting, park2020graphene, yaglikci2021blockhammer, orosa2021deeper, frigo2020trrespass, hassan2021utrr, mutlu2017rowhammer, mutlu2014research, yaglikci2021security, mutlu2019rowhammer}} suggest that many existing proposals may fall short in one or more of these dimensions. As a result, there is a critical need for {developing} better RowHammer mitigation {mechanisms}.}

To {enable} more {effective and efficient} RowHammer {mitigation} {mechanisms}, it is {critical to develop a comprehensive understanding of how RowHammer bit flips occur~\cite{orosa2021deeper, mutlu2017rowhammer, mutlu2019rowhammer}.
In this work, we observe that although the wordline voltage (\gls{vpp}) is expected to affect the amount of disturbance caused by a RowHammer attack~\cite{kim2014flipping, redeker2002investigation, mutlu2017rowhammer, mutlu2019rowhammer, walker2021ondramrowhammer, yang2019trap, park2016statistical, park2016experiments, orosa2021deeper, kim2020revisiting, saroiu2022price, qureshi2021rethinking}, \emph{no} prior work experimentally studies its real-world impact on a DRAM chip's {RowHammer} vulnerability.\footnote{{Both {\gls{vpp}} and {\gls{vdd}} {can affect} a DRAM chip's {RowHammer vulnerability}. However, changing \gls{vdd} can negatively impact DRAM reliability in ways that are unrelated to RowHammer (e.g., I/O circuitry instabilities) because \gls{vdd} supplies power to {\emph{all}} logic elements within the DRAM chip.} In contrast, \gls{vpp} affects \emph{only} the wordline voltage, so \gls{vpp} can influence RowHammer without adverse effects {on unrelated} parts of the DRAM chip.} Therefore,} \textbf{our goal} is to understand {how \gls{vpp} affects} RowHammer vulnerability and DRAM {operation}.  

{To achieve this goal, we start with the hypothesis that \gls{vpp} can be used to reduce a DRAM chip's RowHammer vulnerability without impacting the reliability of normal DRAM operations.} Reducing a DRAM chip's RowHammer vulnerability {via} \gls{vpp} scaling has two key advantages. First, {as a circuit-level RowHammer mitigation approach, \gls{vpp} scaling} is \emph{complementary} to {existing system-level and architecture-level RowHammer mitigation mechanisms~\cite{aichinger2015ddr, AppleRefInc, aweke2016anvil, kim2014flipping, kim2014architectural,son2017making, lee2019twice, you2019mrloc, seyedzadeh2018cbt, van2018guardion, konoth2018zebram, park2020graphene, yaglikci2021blockhammer, kang2020cattwo, bains2015row, bains2016distributed, bains2016row, brasser2017can, gomez2016dummy, jedec2017ddr4,hassan2019crow, devaux2021method, ryu2017overcoming, yang2016suppression, yang2017scanning, gautam2019row, yaglikci2021security, qureshi2021rethinking, greenfield2012throttling, marazzi2022protrr, saileshwar2022randomized}.
Therefore,} \gls{vpp} scaling can be used \emph{alongside} {these mechanisms} to increase their effectiveness and/or reduce their overheads. Second, {\gls{vpp} scaling} can be implemented with a \emph{fixed hardware cost} {for a given power budget,} {irrespective of the number and types of DRAM chips used in a system.}

{We test this hypothesis through the first experimental {RowHammer} characterization study {under reduced \gls{vpp}.}
{In this study, we test} \numchips{} real DDR4 DRAM chips from 
\param{three} major DRAM manufacturers. Our study {is inspired by} state-of-the-art analytical models for RowHammer{, which suggest that the effect of RowHammer's underlying error mechanisms depends on \gls{vpp}~\cite{walker2021ondramrowhammer, yang2019trap, park2016statistical}. \secref{sec:vpp_with_rh} reports our findings, which} yield valuable insights into {how \gls{vpp}} impacts the circuit-level RowHammer characteristics of modern DRAM chips, both confirming our hypothesis and supporting \gls{vpp} scaling as a {promising} new dimension to{ward} robust RowHammer mitigation.}

\section{Experimental Methodology}
\label{sec:methodology}

\agycrone{We \omtwo{describe} our methodology for two analyses. First, we experimentally characterize the behavior of \numchips{} real DDR4 DRAM chips from three major manufacturers under reduced \gls{vpp} in terms of RowHammer vulnerability (\secref{sec:experiment_design_rowhammer}), \acrfull{trcd} (\secref{sec:experiment_design_trcd}), and data retention time (\secref{sec:experiment_design_retention}). Second, to verify our observations from real-device experiments, we investigate reduced \gls{vpp}'s effect on \emph{both} DRAM row activation and charge restoration using SPICE~\cite{ltspice, nagel1973spice} simulations~(\secref{sec:spice_model}).}

\subsection{\agycrone{Real-Device Testing Infrastructure}}
\label{sec:experimental_setup}
\begin{figure}[!ht]
    \centering
    \includegraphics[width=\linewidth]{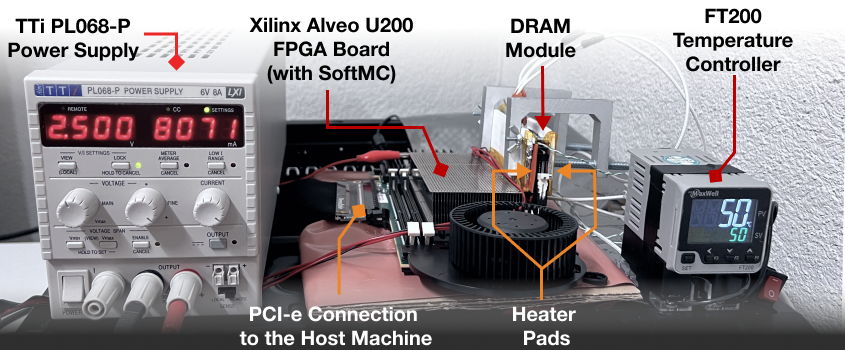}
    \caption{Our experimental setup \omone{based on SoftMC~\cite{hassan2017softmc,softmcgithub}}.}
    \label{fig:infrastructure}
\end{figure}

We conduct \agycrone{real-device} characterization experiments using \omfour{an infrastructure based on} SoftMC~\cite{hassan2017softmc, softmcgithub},\agycrtwocomment{Should we cite DRAMBender instead of SoftMC?} the state-of-the-art FPGA-based open-source \omone{infrastructure for DRAM characterization}. We {extensively} modify SoftMC to test modern DDR4 DRAM chips.
\figref{fig:infrastructure} shows a picture of our experimental setup. {We attach \agycrtwo{\omfour{heater pads}} to the DRAM chips that are located on both sides of a DDR4 DIMM. We use a MaxWell FT200 PID temperature controller\omone{~\cite{maxwellFT200}} connected to the \agycrtwo{heaters pads} to maintain the DRAM chips under test at a preset temperature level with the precision of $\pm$\SI{0.1}{\celsius}.} We program a Xilinx Alveo U200 FPGA board\omone{~\cite{alveo}} with the modified version of SoftMC. The FPGA board is connected to a host machine through a PCI{e} port for running our tests. We connect the DRAM module to {the} FPGA board \omtwo{via} a commercial interposer board from Adexelec\omone{~\cite{adexelecddr4sodimmriser}} with current measurement capability. The interposer board enforces the power to be supplied through a shunt resistor on the \gls{vpp} rail. We remove this shunt resistor to electrically disconnect the \gls{vpp} rails of the DRAM module and the FPGA board. Then, we supply power to the DRAM module's \gls{vpp} power rail from an external TTi PL068-P power supply\agycrone{~\cite{pl068p}}\omone{, which enables us to control \gls{vpp} \agycrone{at the precision of \SI{\pm1}{\milli\volt}.}} \agycrone{We start testing each DRAM module at the nominal \gls{vpp} of \SI{2.5}{\volt}. We gradually reduce \gls{vpp} with \SI{0.1}{\volt} steps until \gls{vppmin}.}

{To show that our observations are 
\emph{not} {specific} to a certain DRAM architecture/process but rather common across different designs and generations, we test {DDR4}} DRAM modules from all three major manufacturers {with different die revisions, purchased from the retail market}.
Table~\ref{tab:dram_chip_list} provides {the} {chip} density, die revision {(Die Rev.)}, chip organization {(Org.)}, and manufacturing date of tested DRAM modules.{\footnote{{Die Rev. and Date columns are blank if undocumented.}}} We report the manufacturing date of these modules in the form of $week-year$.
All tested modules {are listed in Table~\ref{tab:detailed_dimm_table} {in \ref{sec:appendix}}.}

\begin{table}[h]
    \caption{Summary of \omtwo{the tested} DDR4 DRAM chips.}
    \centering
    \footnotesize{}
    \setlength\tabcolsep{3pt} 
    \begin{tabular}{lcccccc}
        \toprule
            {{\bf Mfr.}} & \textbf{\#DIMMs} & {{\bf  \#Chips}}  & {{\bf Density}} & {{\bf Die Rev.}}& {{\bf Org.}}&
            {{\bf Date}}\\
        \midrule
        & 1 & 8 &{4Gb}&{}&{$\times$8}&48-16 \\
        {Mfr. A}& 4 &  64 &{8Gb}&{B}&{$\times$4}&11-19 \\
        \omone{(Micron)}& 3 & 24 &{4Gb}&{F}&{$\times$8}&07-21 \\
        & 2 & 16 &{4Gb}&{}&{$\times$8}&\\
        \midrule
        {}& 2 & 16 &{8Gb}&{B}&{$\times$8}&5\agycrtwo{2}-20\\
        {}& 1 & 8 &{8Gb}&{C}&{$\times$8}&19-19\\
        {Mfr. B}& 3 & 24 &{8Gb}&{D}&{$\times$8}&10-21\\
        \omone{(Samsung)}& 1 &  8 &{4Gb}&{E}&{$\times$8}&08-17\\
        & 1 &  8 &{4Gb}&{F}&{$\times$8}&02-21\\
        & 2 &  16 &{8Gb}&{}&{$\times$8}&\\
        \midrule
        {}& 2 &  16 &{16Gb}&{A}&{$\times$8}&51-20\\
        {Mfr. C}& 3 & 24 &{4Gb}&{B}&{$\times$8}&02-21\\
        \omone{(SK Hynix)}& 2 & 16 &{4Gb}&{C}&{$\times$8}&\\
        & 3 & 24 &{8Gb}&{D}&{$\times$8}&48-20\\

        \bottomrule
    \end{tabular}
    \label{tab:dram_chip_list}
\end{table}

\setstretch{0.90}

\head{Temperature} We conduct RowHammer and \gls{trcd} tests at \SI{50}{\celsius} and retention tests at \SI{80}{\celsius} to ensure both stable and representative testing conditions.{\footnote{A recent work~\cite{orosa2021deeper} shows a complex interaction between RowHammer and temperature{, suggesting {that one should} repeat characterization at many different temperature levels {to find the worst-case RowHammer vulnerability}. Since such characterization requires {many} months-long testing time,} we leave it to future work to study temperature, voltage, and RowHammer interaction in detail.}} 
We conduct \gls{trcd} tests at \SI{50}{\celsius} because \SI{50}{\celsius} is our infrastructure’s minimum stable temperature due to cooling limitations.{\footnote{We do not repeat {the \gls{trcd}} tests at different temperature levels because prior work~\cite{chang2017understanding} shows {small} variation in \gls{trcd} with varying temperature.}} We conduct retention tests at \SI{80}{\celsius} to capture any effects of increased charge leakage~\cite{liu2013experimental} {at the upper bound of regular operating temperatures~\cite{jedec2017ddr4}.\footnote{DDR4 DRAM chips are refreshed at $2\times$ the nominal refresh rate when the chip temperature reaches \SI{85}{\celsius}~\cite{jedec2017ddr4}. Thus, we choose \SI{80}{\celsius} as a representative high temperature within the regular operating temperature range. {For a detailed analysis of the effect of temperature on data retention in DRAM, we refer the reader to~\cite{hamamoto1998on,liu2013experimental, patel2017reaper}.}}}

\head{Disabling Sources of Interference} 
{To understand fundamental device behavior in response to \gls{vpp} reduction, we} make sure that \gls{vpp} is the only {control} variable in our experiments {so that we can} {accurately} measure the effects of \gls{vpp} on 
{RowHammer, row activation latency (\gls{trcd}), and data retention time.}
To do so, {we follow {four} steps{, similar to prior rigorous RowHammer~\cite{kim2020revisiting,orosa2021deeper}, row activation latency~{\cite{lee2015adaptive, chang2016understanding, chang2017understanding}}, and data retention time~\cite{liu2013experimental, patel2017reaper} characterization methods}. F}irst{,} we disable DRAM refresh {to ensure no disturbance on the desired access pattern.}
Second, we ensure that during our {RowHammer and \gls{trcd}} experiments{,} {\emph{no}} bit flips occur due to {data} retention failures by conducting {each experiment} within a {time} period {of less than  \SI{30}{\milli\second}} {(i.e., much shorter than the nominal \gls{trefw} of \SI{64}{\milli\second})}. Third, we test DRAM modules 
{without} error{-}correction code (ECC) support to ensure neither on-die {ECC}~{\cite{ECCMicron,nair2016xed,patel2021harp,patel2020beer,patel2019understanding,kang2014co,mineshphd}} nor rank-level ECC~\cite{cojocar2019eccploit,kim2016all} can affect {our observations} by correcting {\gls{vpp}-reduction-induced} bit flips.
{Fourth}, we \omone{disable} known on-DRAM-die RowHammer
defenses (i.e., TRR~{\cite{jedec2020lpddr5,jedec2020ddr5,lee2014green,micron2016trr,hassan2021utrr, frigo2020trrespass}}) by not issuing refresh commands throughout our tests~\cite{frigo2020trrespass,kim2020revisiting,orosa2021deeper, hassan2021utrr} {(as all {TRR} defenses require refresh commands to work)}. 

\head{{Data Patterns}} {We use} six commonly used data patterns~\cite{chang2016understanding,chang2017understanding,khan2014efficacy,khan2016parbor,khan2016case,kim2020revisiting,lee2017design,mukhanov2020dstress,orosa2021deeper, kim2014flipping, liu2013experimental}: row stripe (\texttt{0xFF}/\texttt{0x00}), checkerboard (\texttt{0xAA}/\texttt{0x55}), and thickchecker (\texttt{0xCC}/\texttt{0x33}). We identify {the worst-case data pattern ($WCDP$) for each row among these {six patterns} at nominal \gls{vpp} {separately} {for each of RowHammer (\secref{sec:experiment_design_rowhammer}), \acrfull{trcd} (\secref{sec:experiment_design_trcd}), and data retention time (\secref{sec:experiment_design_retention}) tests. We {use} each row{'s}
{corresponding} $WCDP$ {for a given test,} at reduced \gls{vpp} levels.}}

\setstretch{0.9225}
\subsection{\omone{RowHammer Experiments}}
\label{sec:experiment_design_rowhammer}
{We perform multiple experiments to understand how \gls{vpp} affects the Row{H}ammer vulnerability of a DRAM chip.}

\head{Metrics}~We measure the RowHammer vulnerability of a DRAM chip {using} two {metrics}: {1)
\acrlong{hcfirst} (\acrshort{hcfirst}) and 2)~\gls{ber} {{caused} by} a double-sided RowHammer attack with a fixed hammer count of 300K per aggressor row.}{\footnote{{{We choose the 300K hammer count because 1) it is low enough to be used in a system-level RowHammer attack in a real system, and 2) it is high enough to provide us with a large number of bit flips to make meaningful observations in all DRAM modules we tested.}}}}

\head{\agycrone{WCDP}}
{We choose $WCDP$ as the data pattern {that} causes the \emph{lowest} \gls{hcfirst}. If there are multiple data patterns that {cause} the lowest \gls{hcfirst}, we choose the data pattern that causes the \emph{largest} \gls{ber} for the fixed hammer count of $300K$.}\footnote{{To investigate if $WCDP$ changes with reduced \gls{vpp},} {we repeat {$WCDP$ determination} experiments for different \gls{vpp} values for \param{16} DRAM chips.} We observe that $WCDP$ changes for \emph{only} \SI{2.4}{\percent} of tested rows,  causing less than \SI{9}{\percent} deviation in \gls{hcfirst} for \SI{90}{\percent} of the affected rows. We leave a {detailed sensitivity analysis of $WCDP$ to \gls{vpp}}
for future work.}

{\head{RowHammer Tests} Alg.~\ref{alg:test_alg} describes the core test loop of each RowHammer test that we run. The algorithm performs a \emph{double-sided} RowHammer {attack} on each row within a DRAM bank. {A double-sided RowHammer attack activates the two attacker rows that are physically adjacent to a victim row (i.e., the victim row's two immediate neighbors) in an alternating manner. 
{We define \gls{hc} as the number of times each
physically-adjacent row is activated.}
In this study, we perform double-sided attacks instead of {single-}~\cite{kim2014flipping} or many-sided attacks (e.g., {as in} TRRespass~\cite{frigo2020trrespass}, U-TRR~\cite{hassan2021utrr}, and BlackSmith~\cite{jattke2022blacksmith}) because a double-sided attack is the most effective RowHammer attack when no RowHammer defense mechanism is employed{:} {it reduces \gls{hcfirst} and increases \gls{ber} compared to {both} single- {and many-}sided attack{s}}~\cite{orosa2021deeper, kim2014flipping, kim2020revisiting, frigo2020trrespass, hassan2021utrr, jattke2022blacksmith}. 
{Due to time limitations, 1)~we test $4K$ rows {per DRAM module} (four chunks of $1K$ rows evenly distributed across a DRAM bank) and 2)~we run each test ten times and record the smallest (largest) observed \gls{hcfirst} (\gls{ber}) to account for the worst-case.}
}}

\SetAlFnt{\scriptsize}
\RestyleAlgo{ruled}
\begin{algorithm}
\caption{\omtwo{Test for} \gls{hcfirst} and \gls{ber} \omthree{for a Given} \gls{vpp}}\label{alg:test_alg}
  \DontPrintSemicolon
  \SetKwFunction{FVPP}{set\_vpp}
  \SetKwFunction{FMain}{test\_loop}
  \SetKwFunction{FHammer}{measure\_$BER$}
  \SetKwFunction{initialize}{initialize\_row}
  \SetKwFunction{initializeaggr}{initialize\_aggressor\_rows}
  \SetKwFunction{measureber}{measure\_BER}
  \SetKwFunction{FMeasureHCfirst}{measure\_$HC_{first}$}
  \SetKwFunction{compare}{compare\_data}
  \SetKwFunction{Hammer}{hammer\_doublesided}
  \SetKwFunction{Gaggressors}{get\_aggressors}
  \SetKwFunction{FWCDP}{get\_WCDP}
  \SetKwProg{Fn}{Function}{:}{}

  \tcp{$RA_{victim}$: victim row address}
  \tcp{$WCDP$: worst-case data pattern}
  \tcp{$HC$: number \omone{of} activations per aggressor row}
  \Fn{\FHammer{$RA_{victim}$, $WCDP$, $HC$}}{
        \initialize($RA_{victim}$, $WCDP$)\;
        \omone{\initializeaggr($RA_{victim}$, \agycrtwo{bitwise\_inverse($WCDP$)})}\;
        \Hammer(\omone{$RA_{victim}$}, $HC$)\;
        $BER\omone{_{row}} =$ \compare($RA_{victim}$, $WCDP$)\;
        \KwRet $BER\omone{_{row}}$\;
  }\;
  
  \tcp{$V_{pp}$: \omone{wordline voltage for the experiment}}
  \tcp{\omone{$WCDP\_list$: the list of $WCDP$s (one $WCDP$ per row)}}
  \tcp{\omone{$row\_list$: the list of tested rows}}
  \Fn{\FMain{$V_{pp}$, \omone{$WCDP$\_list}}}{
        \FVPP($V_{pp}$)\;
        \ForEach{$RA_{victim}$ in $row\_list$}{
            \omone{$HC = 300K$ // initial hammer count to test\;
            \omone{$HC_{step} = 150K$} // how much to increment/decrement $HC$ \;
            \While{$HC_{step} > 100$}{
                $BER_{row_{max}} = 0$\;
                \For{$i\gets0$ \KwTo $num\_iterations$}{ 
                    \agycrone{$BER_{row} = ~$\measureber($RA_{victim}$, $WCDP$, $HC$)}\;
                    record\_BER($V_{pp}$, $RA_{victim}$, $WCDP$, $HC$, $BER_{row}$, $i$)\;
                    $BER_{row_{max}} =  max(BER_{row_{max}}, BER_{row})$\;
                }
                \If{$BER_{row_{max}} == 0$}{
                  \agycrtwo{$HC += HC_{step}$ // Increase HC if no bit flips occur}
                }\Else{
                  \agycrthree{$HC -= HC_{step}$ // Reduce HC if a bit flip occurs}
                }
                $HC_{step}$ = $HC_{step}/2$\;
            }
            record\_HCfirst($V_{pp}$, $RA_{victim}$, $WCDP$, $HC$)\;
        }}

  }\;
\end{algorithm}

\noindent
{\textbf{Finding Physically Adjacent Rows.} DRAM-internal address mapping schemes~\cite{cojocar2020rowhammer,salp} are used by DRAM manufacturers to translate {\emph{logical}} DRAM addresses (e.g., row, bank, {and} column) that are exposed over the DRAM interface (to the memory controller) to physical {DRAM} addresses {(e.g., physical location of a row)}. {Internal address mapping schemes allow 
{1)~}post-manufacturing row repair techniques to repair erroneous DRAM rows by remapping these rows to spare rows and 
{2)~}DRAM manufacturers to organize DRAM internals in a cost-optimized way, e.g., by organizing internal DRAM buffers hierarchically~\cite{khan2016parbor,vandegoor2002address}.} The mapping scheme can vary {substantially} across different DRAM {chips}~\cite{barenghi2018software,cojocar2020rowhammer,horiguchi1997redundancy,itoh2013vlsi,keeth2001dram,khan2016parbor,khan2017detecting,kim2014flipping,lee2017design,liu2013experimental,patel2020beer,orosa2021deeper,saroiu2022price,patel2022case}. For every victim DRAM row {we test}, we identify the two {neighboring physically-adjacent} DRAM row addresses that the memory controller can use to access the {aggressor} rows in a double-sided RowHammer attack. To do so, we} reverse-engineer the physical row organization {using} techniques described in prior work{s}~\cite{kim2020revisiting, orosa2021deeper}.

\subsection{{Row Activation Latency (\gls{trcd}) Experiments}}
\label{sec:experiment_design_trcd}
{{We conduct experiments t}o find how a DRAM chip's {row activation latency (}\gls{trcd}{)} changes with {reduced} \gls{vpp}.}

\head{Metric} We {measure} \gls{trcdmin} {between a row activation and the {following} read operation {to ensure {that} there are {\emph{no}}} bit flips in the entire DRAM row}.

\head{\omone{WCDP}}
{We choose $WCDP$ as the data pattern {that} {leads to} the \emph{largest} {observed} \gls{trcdmin}.}

\noindent\textbf{$\mathbf{t_{RCD}}$ Tests.} Alg.~\ref{alg:trcd_test_alg} describes {the core test loop of each \gls{trcd} test that we run.}
The algorithm sweeps \gls{trcd} starting from {the nominal \gls{trcd} of \SI{13.5}{\nano\second} with steps of} \SI{1.5}{\nano\second}.\footnote{{Our version of SoftMC can send a DRAM command every \SI{1.5}{\nano\second} due to the clock frequency limitations in the FPGA's physical {DRAM} interface.}} {We decrement (increment) \gls{trcd} by \SI{1.5}{\nano\second} until we observe at least one (no) bit flip in the entire DRAM row}
in order to {pinpoint} \gls{trcdmin}. 
To {test a DRAM row for a given \gls{trcd}, the algorithm} 1)~initializes the {row with {the row's}} $WCDP$, 2)~performs an access {using the given \gls{trcd} for each column in the row}
and 3)~checks if the access results in any {bit flips}. 
{After {testing each} column in a DRAM row, the algorithm identifies
{the row's \gls{trcdmin} as the minimum \gls{trcd} that does not cause any bit flip in the entire DRAM row.}
{{Due to time limitations, we 1)~test the same set of rows as we use in RowHammer tests {(\secref{sec:experiment_design_rowhammer})} and 2)~run each test ten times}
and record the {\emph{largest} \gls{trcdmin} for each row} across all runs.}}\footnote{{To understand whether {reliable} DRAM row activation latency changes over time, we repeat these tests for 24 DRAM chips after one week, during which the chips are tested for RowHammer vulnerability. We observe that \emph{only} {\SI{2.1}{\percent}} of tested DRAM rows experience only a small variation ($<$\SI{1.5}{\nano\second}) in \gls{trcd}. {This result is consistent with results of prior works~\cite{chang2016understanding, chang2017understanding,kim2018solar}.}}}

\SetAlFnt{\scriptsize}
\RestyleAlgo{ruled}
\begin{algorithm}
\caption{Test {for} {Row} Activation Latency {for a Given} \gls{vpp}}\label{alg:trcd_test_alg}
  \DontPrintSemicolon
  \SetKwFunction{FVPP}{set\_vpp}
  \SetKwFunction{FMain}{test\_loop}
  \SetKwFunction{Ftrcd}{identify\_\gls{trcdmin}}
  \SetKwFunction{compare}{compare}
  \SetKwFunction{initialize}{initialize\_row}
  \SetKwFunction{initializeneighbor}{initialize\_neighboring\_rows}
  \SetKwFunction{FWCDP}{get\_WCDP}
  \SetKwProg{Fn}{Function}{:}{}
  
  \tcp{$V_{pp}$: {wordline voltage for the experiment}}
  \tcp{{$WCDP\_list$: the list of WCDPs (one WCDP per row)}}
  \tcp{{$row\_list$: the list of tested rows}}
    \Fn{\FMain{$V_{pp}$, {$WCDP\_list$, $row\_list$}}}{
    \FVPP($V_{pp}$)\;        
      \ForEach{{$RA$ in $row\_list$}}{
        \gls{trcd} = {\SI{13.5}{\nano\second}}\;
        {found\_faulty, found\_reliable = False, False\;}
        \While{{not found\_faulty or not found\_reliable}}
        {
            {is\_faulty = False\;}
            \For{{$i\gets0$ \KwTo $num\_iterations$}}{
                \ForEach{{column $C$ in row $RA$}}{
                    {\initialize($RA$, $WCDP\_list[\omone{RA}]$)}\;
                    {activate\_row($RA$, \gls{trcd}) //activate the row using \gls{trcd}}\;
                    {read\_data = read\_col($C$)}\;
                    {close\_row($RA$)}\;
                    {$BER_{col}$ = \compare(WCDP\_list[\omone{RA}], read\_data)}\;
                    {\lIf{$BER_{col}$ > 0}{is\_faulty=True}}
                }
            }
            {\lIf{is\_faulty}{
                \{\gls{trcd} += \SI{1.5}{\nano\second};
                found\_faulty = True;\}
            }
            \lElse{
                \{\gls{trcdmin} = \gls{trcd};
                \gls{trcd} -= \SI{1.5}{\nano\second};
                found\_reliable = True;\}
            }
        }
        }
        
        {record\_\gls{trcdmin}($RA$, \gls{trcdmin})}\;
    }
  }
\end{algorithm}

\subsection{{Data Retention Time Experiments}}
\label{sec:experiment_design_retention}
{We conduct {data} retention time experiments t}{o understand the effects of \gls{vpp} on DRAM cell data retention characteristics. 
{We test the same set of DRAM rows as we use in RowHammer tests {(\secref{sec:experiment_design_rowhammer})} for a set of fixed refresh windows from \SI{16}{\milli\second} to {\SI{16}{\second}} in increasing powers of two.}}

\head{Metric} We measure {\acrlong{ber} ({retention-}\gls{ber})} {due to violating a DRAM row's data retention time, using a reduced refresh rate.}

\head{{WCDP}} {We choose $WCDP$ as the data pattern which causes a bit flip at the \emph{smallest} refresh window (\gls{trefw}) among the six data patterns. If we find more than one such data {pattern}, we choose the one that {leads to} the largest \gls{ber} for \gls{trefw} of \SI{16}{\second}.}

\noindent
{\textbf{{Data} Retention Time Tests.} Alg.~\ref{alg:ref_test_alg} describes how we perform {data} retention tests to {measure} {retention-}\gls{ber} for a given \gls{vpp} and {refresh rate}.
{The algorithm} 1)~initialize{s} a DRAM row with {WCDP}, 2)~waits {as long as the given refresh window}, and {3})~read{s} and compare{s} the data in the DRAM row {to the row's initial data}{.}} 

\SetAlFnt{\scriptsize}
\RestyleAlgo{ruled}
\begin{algorithm}
\caption{Test {for} {Data} Retention Time{s} {for a Given} \gls{vpp}}\label{alg:ref_test_alg}
  \DontPrintSemicolon
  \SetKwFunction{FVPP}{set\_vpp}
  \SetKwFunction{FMain}{test\_loop}
  \SetKwFunction{Fref}{measure\_ber}
  \SetKwFunction{access}{accessDRAM}
  \SetKwFunction{initialize}{initialize\_row}
  \SetKwFunction{initializeneighbor}{initialize\_neighboring\_rows}
  \SetKwFunction{compare}{compare\_data}
  \SetKwFunction{wait}{wait}
  \SetKwFunction{FWCDP}{get\_WCDP}
  \SetKwProg{Fn}{Function}{:}{}
  
  \tcp{$V_{pp}$: {wordline voltage for the experiment}}
  \tcp{{$WCDP\_list$: the list of WCDPs (one WCDP per row)}}
  \tcp{{$row\_list$: the list of tested rows}}
  \Fn{\FMain{$V_{pp}$, {$WCDP\_list$, $row\_list$}}}{
      \FVPP($V_{pp}$)\;
      \gls{trefw} = {\SI{16}{\milli\second}}\;
      \While{{\gls{trefw} $\leq$ \SI{16}{\second}}}{
        \For{{$i\gets0$ \KwTo $num\_iterations$}}{
            \ForEach{{$RA$ in $row\_list$}}{
            {\initialize($RA$, $WCDP\_list[{RA}]$)}\;      
            {wait(\gls{trefw})}\;
            {read\_data = read\_row($RA$)}\;
            {$BER_{row}$ = \compare(WCDP\_list[{RA}], read\_data)}\;
            {record\_retention\_errors($RA$, \gls{trefw}, $BER_{row}$)}\;
            }
        }
        \gls{trefw} = \gls{trefw}$\times2$\;
      }
      
  }
\end{algorithm}

\subsection{SPICE Model}
\label{sec:spice_model}
To {provide insights into} our {real-chip-based} experimental observations {about}
{the effect of reduced \gls{vpp} on row activation latency and data retention time,}
we conduct a set of SPICE~\cite{ltspice, nagel1973spice} simulations {to estimate the bitline and cell voltage levels} {during two relevant DRAM operations: row activation and charge restoration}. To do so, we adopt {and modify} a SPICE model used in a relevant prior work~\cite{chang2017understanding} that studies the impact of changing \gls{vdd} (but \emph{not} \gls{vpp}) on DRAM row access and refresh operations.
{Table~\ref{tab:spice-param} summarizes our SPICE model, which we open-source~\cite{rowhammervppgithub}.}
We use LTspice~\cite{ltspice} with the \SI{22}{\nano\meter} PTM transistor model~\cite{ptmweb, zhaoptm} {and scale} the simulation parameters
{according to the} ITRS {roadmap}~\cite{itrs_model,vogelsang2010understanding}.\footnote{{We} do \emph{not} expect {SPICE simulation and real-world experimental results to be identical because {a} SPICE model \emph{cannot} simulate a real DRAM chip's exact behavior without}
proprietary design and manufacturing information.} {To account for manufacturing process variation, }{we perform Monte-Carlo {simulations} by randomly varying the component parameters up to \SI{5}{\percent} for each simulation run. {We} run the simulation {at \gls{vpp} levels from \SI{1.5}{\volt} to \SI{2.5}{\volt} with a step size of at \SI{0.1}{\volt}} {10K} times, {similar to prior works~\cite{hassan2019crow, luo2020clrdram}}. 
}

\begin{table}[h!]
\caption{Key parameters used in SPICE {simulations}.
}
\footnotesize
\centering
\begin{tabular}{ll}
\toprule
\textbf{Component} & \textbf{Parameters}  \\
\midrule
DRAM Cell          & C: 16.8 fF, R: 698 $\Omega$  \\

Bitline            & C: 100.5 fF, R: 6980 $\Omega$\\

Cell Access NMOS   & W: 55 nm, L: 85 nm    \\

Sense Amp. NMOS    & W: 1.3 um, L: 0.1 um \\

Sense Amp. PMOS    & W: 0.9 um, L: 0.1 um \\
\bottomrule
\end{tabular}
\label{tab:spice-param}
\end{table}

\subsection{{Statistical Significance of Experimental Results}}
\label{sec:statistical:sig}

To evaluate the statistical significance of our methodology, we investigate the variation in our measurements by examining the \emph{coefficient of variation (CV)}
across ten iterations. {CV is a standardized metric to measure the extent of variability in a set of measurements, in relation to the mean of the measurements. CV is calculated as the ratio of standard deviation over the mean value~\cite{Everitt1998cambridgestatistics}. A smaller CV shows a smaller variation across measurements, indicating {higher} statistical significance.} 
The coefficient of variation is 0.08, 0.13, and 0.24 for $90^{th}$, $95^{th}$, and $99^{th}$ {percentiles} of {all of our experimental results}, respectively.

\section{RowHammer Under Reduced \gls{vpp}}
\label{sec:vpp_with_rh}

We provide the first experimental characterization of how {\acrfull{vpp}} affects the RowHammer vulnerability 
of {a} DRAM row in terms of 1)~\acrfull{ber} {(\secref{sec:vpp:ber})} and 2)~\acrfull{hcfirst} {(\secref{sec:vpp_vs_hcfirst}). To conduct this analysis{,} we provide experimental results {from \numchips{}} real DRAM {chips}, {using} the methodology described in \secref{sec:experimental_setup} and \secref{sec:experiment_design_rowhammer}.}

\subsection{{Effect} of \gls{vpp} on {RowHammer BER}}
\label{sec:vpp:ber}
\begin{figure}[!ht]
    \centering
    \includegraphics[width=\linewidth]{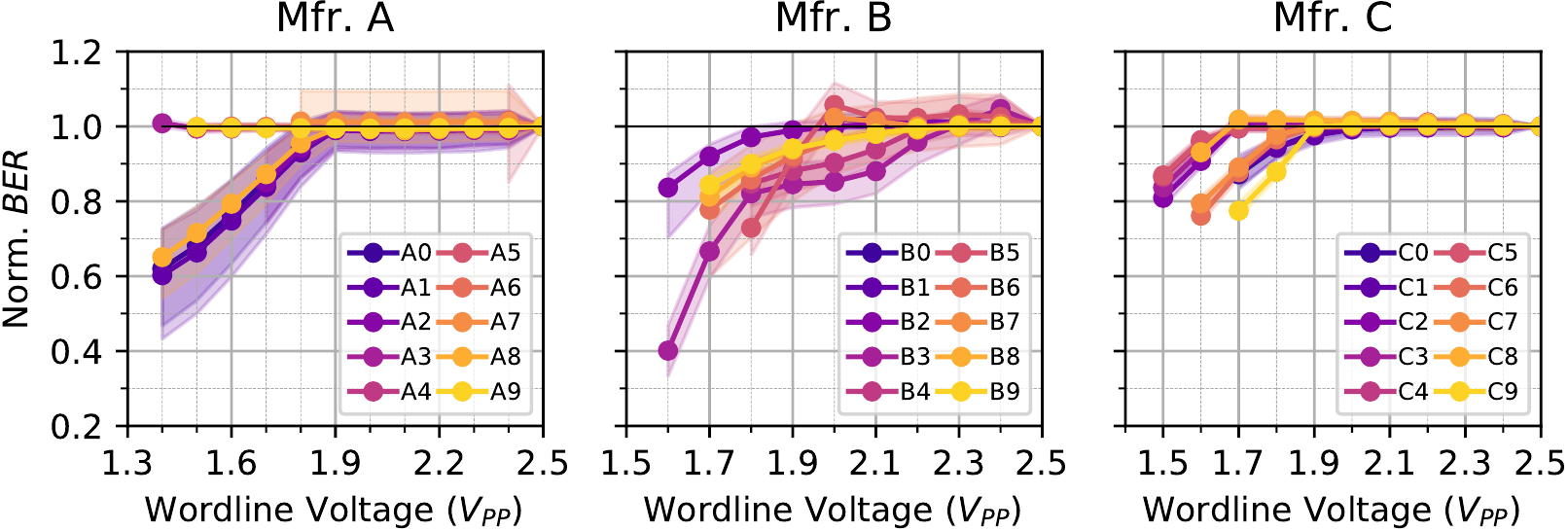}
    \caption{Normalized \gls{ber} values across different \gls{vpp} levels. Each curve represents a different DRAM module.}
    \label{fig:vpp_ber}
\end{figure}

\noindent
\figref{fig:vpp_ber} shows the {RowHammer} \gls{ber} a DRAM row experiences at a fixed hammer count of $300K$ under different voltage levels, normalized to the row's {RowHammer} \gls{ber} at nominal \gls{vpp} (\SI{2.5}{\volt}). Each line represents a different DRAM module. The band of shade around each line marks the {\SI{90}{\percent}} confidence interval of the normalized \gls{ber} value across all tested DRAM rows. We make \obsvsref{obsv:dominant_ber} and~\ref{obsv:discrepency_ber} from \figref{fig:vpp_ber}.

\observation{Fewer DRAM cells experience bit flips due to {RowHammer} under reduced wordline voltage. \label{obsv:dominant_ber}}

We observe that {RowHammer} \gls{ber} \emph{decreases} as \gls{vpp} reduces {in \fracbersupportingrows{} of tested rows across all tested modules}. This reduction in \gls{ber} reaches up to \berdecrmax{} (B3 {at $V_{PP}=1.6V$}) with an average of \berdecravg{} {(not shown in the figure)} across all modules we test. {We} conclude that the disturbance caused by hammering a DRAM row becomes weaker{, on average,} with reduc{ed} \gls{vpp}. 

\observation{In contrast to the dominant trend, reducing \gls{vpp} can {sometimes} increase \gls{ber}. \label{obsv:discrepency_ber}}

{We observe that \gls{ber} {increases in \fracberopposingrows{} of tested rows} with reduced \gls{vpp} {by up to {\berincrmax{}}}
{(B5 at $V_{PP} = 2.0V$)}.
We {suspect that {the} {\gls{ber} increase} {we observe} 
occurs {due to a} weakened charge restoration process rather than an {actual} increase in {read} disturbance {(due to} RowHammer{)}.} 
{\secref{sec:sideeffects_retention}} analyze{s} the impact of reduc{ed} \gls{vpp} on {the} charge restoration process.} 

\begin{figure}[!h]
    \centering
    \includegraphics[width=\linewidth]{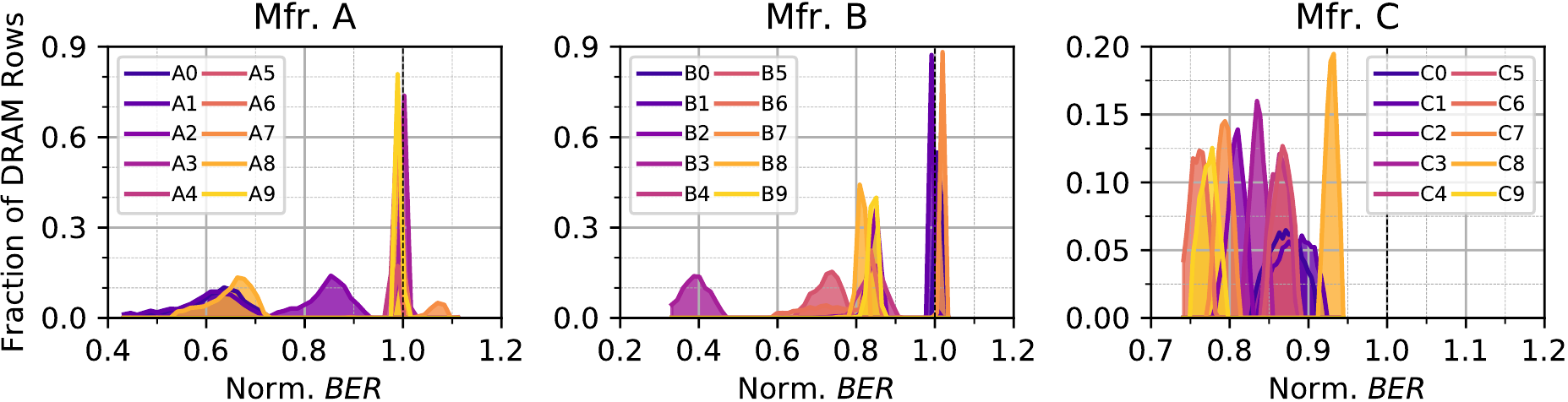}
    \caption{Population density distribution of DRAM rows based on their {normalized} \gls{ber} values at \gls{vppmin}.}
    \label{fig:vpp_ber_fine}
\end{figure}

\head{{Variation in \gls{ber} Reduction Across DRAM Rows}}
{We} investigate how \gls{ber} {reduction {with reduced \gls{vpp}} varies across DRAM rows. To do so, we measure \gls{ber} reduction of each DRAM row at \gls{vppmin} {(\secref{sec:experimental_setup})}.}
\figref{fig:vpp_ber_fine} shows a population density distribution of DRAM rows {(y-axis) based on their \gls{ber} at \gls{vppmin}, normalized to their \gls{ber} at {the} {nominal \gls{vpp} level} (x-axis),} for each manufacturer.
We make \obsvref{obsv:hist_ber} from \figref{fig:vpp_ber_fine}.

\observation{{\gls{ber} reduction with reduced \gls{vpp} varies across different DRAM rows and different manufacturers.}\label{obsv:hist_ber}}

{DRAM rows exhibit a large range of normalized \gls{ber} values {(\minnormbermfrA{}--\maxnormbermfrA{}, \minnormbermfrB{}--\maxnormbermfrB{}, and \minnormbermfrC{}--\maxnormbermfrC{} in chips from Mfrs. A, B, and C, respectively)}.}
{\gls{ber} reduction also varies across different manufacturers. For example, \gls{ber} reduces by more than \SI{5}{\percent} for \emph{all} DRAM rows {of} Mfr.~C, while \gls{ber} variation with reduced \gls{vpp} is smaller than \bersmallchange{} in \fracbersmallchangerowsmfrA{} of the rows {of} Mfr.~A.}

Based on \obsvsref{obsv:dominant_ber}--\ref{obsv:hist_ber}, we conclude that {a DRAM row's} RowHammer \gls{ber} {tends to decrease with reduced \gls{vpp}, while both the amount and the direction of change in \gls{ber} varies across different DRAM rows and manufacturers.}

\subsection{{Effect} of \gls{vpp} on \gls{hcfirst}}
\label{sec:vpp_vs_hcfirst}

\begin{figure}[!ht]
    \centering
    \includegraphics[width=\linewidth]{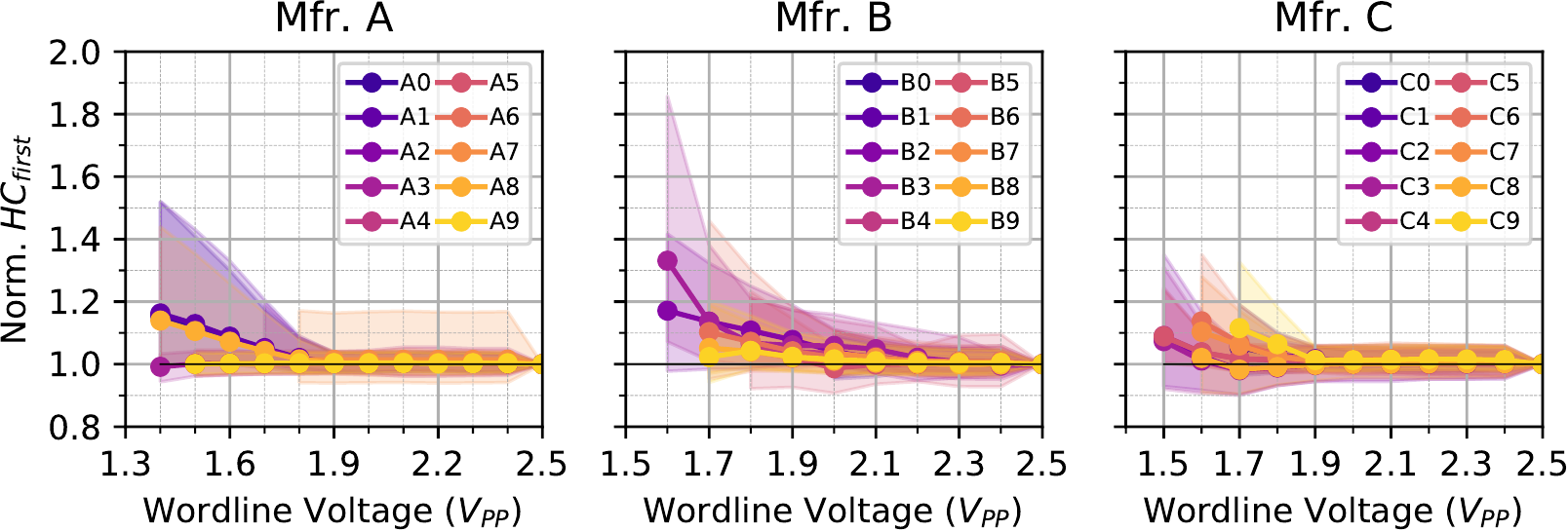}
    \caption{Normalized \gls{hcfirst} values across different \gls{vpp} levels. Each curve represents a different DRAM module.}
    \label{fig:vpp_hcfirst}
\end{figure}

\figref{fig:vpp_hcfirst} shows the \gls{hcfirst} a DRAM row exhibits under different voltage levels, normalized to the row's \gls{hcfirst} at nominal \gls{vpp} (\SI{2.5}{\volt}). Each line represents a different DRAM module. The band of shade around each line marks the {\SI{90}{\percent}} confidence interval of the normalized \gls{hcfirst} value{s} across all tested DRAM rows {in the module}. We make \obsvsref{obsv:dominant_hcfirst} and~\ref{obsv:discrepency_hcfirst} from \figref{fig:vpp_ber}.

\observation{DRAM cells experience RowHammer bit flips at higher hammer counts under reduced wordline voltage. \label{obsv:dominant_hcfirst}}

We observe that \gls{hcfirst} of a DRAM row increases as \gls{vpp} reduces {in} \frachcfirstsupportingrows{} of {tested rows across all tested modules.} This increase in \gls{hcfirst} reaches up to \hcfirstincrmax{} (B3 {at $V_{PP}=1.6V$}) with an average of \hcfirstincravg{} {(not shown in the figure)} across all {tested} modules. {W}e conclude that the disturbance caused by hammering a DRAM row becomes weaker with reduc{ed} \gls{vpp}. 

\observation{In contrast to the dominant trend, reducing \gls{vpp} can {sometimes} cause bit flips at lower hammer counts. \label{obsv:discrepency_hcfirst}}

We observe that \gls{hcfirst} 
reduces {in \frachcfirstopposingrows{} of tested rows with reduced \gls{vpp} by up to \hcfirstdecrmax{} (C8 at \gls{vpp}=\SI{1.6}{\volt})}. {Similar to} \obsvref{obsv:discrepency_ber}, {we suspect that this behavior is caused by the weakened charge restoration process} {(see \secref{sec:sideeffects_retention})}. 
 
\begin{figure}[!ht]
    \centering
    \includegraphics[width=\linewidth]{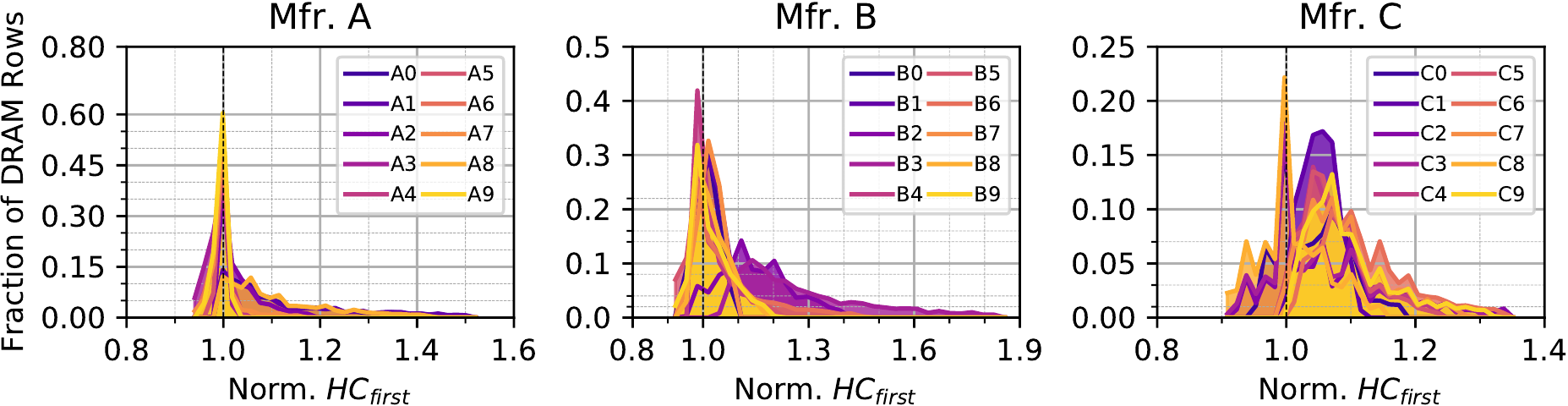}
    \caption{Population density distribution of DRAM rows based on their {normalized} \gls{hcfirst} values at \gls{vppmin}.}
    \label{fig:vpp_hcfirst_fine}
\end{figure}

\head{{Variation in \gls{hcfirst} Increase Across DRAM Rows}}
{We} investigate how \gls{hcfirst} {increase varies {with reduced \gls{vpp}} across DRAM rows. To do so, we measure \gls{hcfirst} {increase} of each DRAM row at \gls{vppmin} {(\secref{sec:experimental_setup})}.}
\figref{fig:vpp_hcfirst_fine} shows a population density distribution of DRAM rows {(y-axis) based on their \gls{hcfirst} at \gls{vppmin}, normalized to their \gls{hcfirst} at {the} {nominal \gls{vpp} level} (x-axis),} for each manufacturer.
We make \obsvref{obsv:hist_hcfirst} from \figref{fig:vpp_hcfirst_fine}.

\observation{{\gls{hcfirst} increase with reduced \gls{vpp} varies across different DRAM rows and different manufacturers.}\label{obsv:hist_hcfirst}}

{DRAM rows in chips from the same manufacturer exhibit a large range of normalized \gls{hcfirst} values (\minnormhcfirstmfrA{}--\maxnormhcfirstmfrA{}, \minnormhcfirstmfrB{}--\maxnormhcfirstmfrB{}, and \minnormhcfirstmfrC{}--\maxnormhcfirstmfrC{} for Mfrs. A, B, and C, respectively).}
{\gls{hcfirst} increase {also} varies across different manufacturers. For example, \gls{hcfirst} increases with reduced \gls{vpp} for
{\frachcfirstsupportingrowsmfrC{} of DRAM rows in modules from Mfr.~C, while \frachcfirstsupportingrowsmfrA{} of DRAM rows exhibit this behavior in modules from Mfr.~A.}
}

Based on \obsvsref{obsv:dominant_hcfirst}--\ref{obsv:hist_hcfirst}, we conclude that {a DRAM row's \gls{hcfirst} tends to increase with reduced \gls{vpp}, while both the amount and the direction of change in \gls{hcfirst} varies across different DRAM rows and manufacturers.}

{\head{{Summary of Findings}}
Based on our analyses on both \gls{ber} and \gls{hcfirst}, we conclude that {a DRAM chip's RowHammer vulnerability can be reduced}
by operating the chip at a \gls{vpp} level {that} is lower than the nominal \gls{vpp} value{.}}

\section{DRAM \agycrone{Reliability} Under Reduced \gls{vpp}}
\label{sec:sideeffects}

{To investigate the effect of reduced \gls{vpp} on {reliable} DRAM {operation},}
we provide the first experimental characterization of how \gls{vpp} affects the {reliability} of three \gls{vpp}-related fundamental DRAM operations: 1)~DRAM row activation {(\secref{sec:sideeffects_trcd})}, 2)~charge restoration {(\secref{sec:sideeffects_charge_restoration})}, and~3)~DRAM refresh {(\secref{sec:sideeffects_retention})}. 
To conduct these analyses{,} we provide both {1)} experimental results {from} real DRAM devices, {using} the methodology described in {\secref{sec:experimental_setup}, \secref{sec:experiment_design_trcd}, and \secref{sec:experiment_design_retention}} and 2) SPICE simulation results{,} {using} the methodology described in {\secref{sec:spice_model}}. 

\subsection{DRAM Row Activation Under Reduced \gls{vpp}}
\label{sec:sideeffects_trcd}

\noindent
\textbf{Motivation.} 
{DRAM row activation latency (\gls{trcd}) should theoretically increase with reduced \gls{vpp} (\secref{sec:background:dramvoltage}).}
{We} {investigate} how \gls{trcd} of real DRAM chips change with {reduced} \gls{vpp}.   

\noindent
\textbf{Novelty.}~{We provide the first experimental analysis of} the isolated impact of \gls{vpp} on activation latency. Prior work~\cite{chang2017understanding} tests DDR3 DRAM chips under reduced {supply voltage (}\gls{vdd}{)}, which may or may not change internally{-}generated \gls{vpp} level. In contrast, we modify only wordline voltage (\gls{vpp}) without {modifying \gls{vdd} to avoid {the possibility} of negatively impacting DRAM reliability due to I/O circuitry instabilities (\secref{sec:background:dramvoltage}).}

\begin{figure}[!ht]
    \centering
    \includegraphics[width=\linewidth]{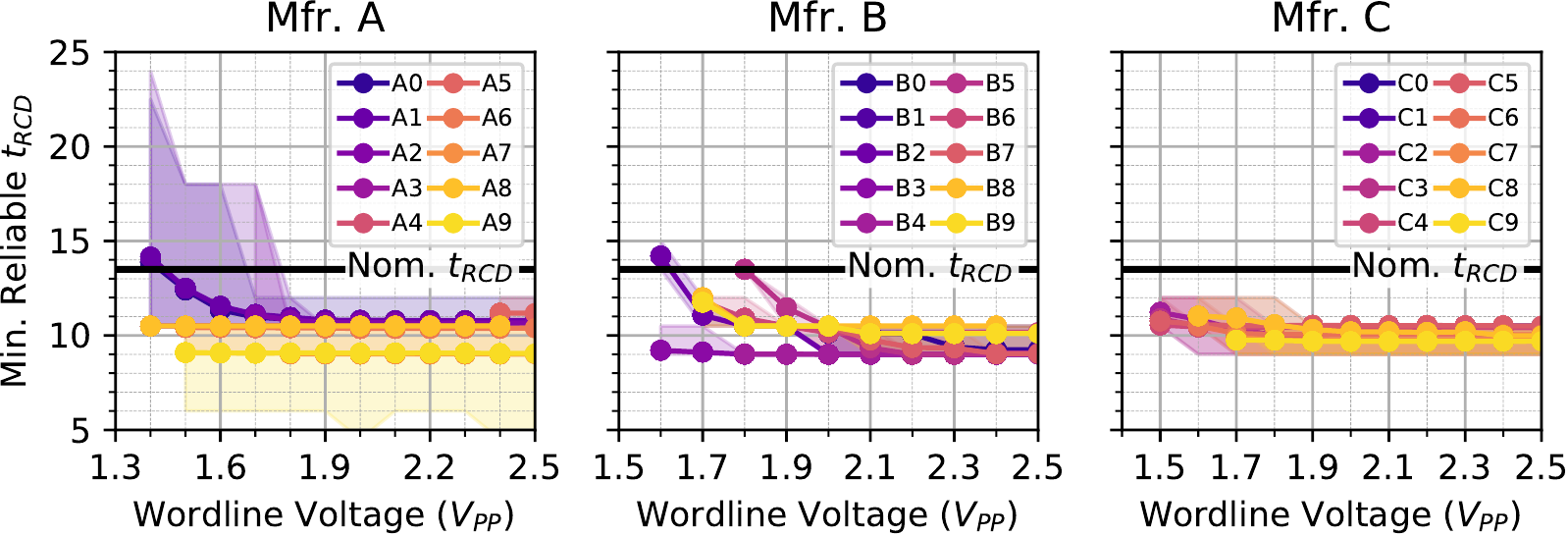}
    \caption{Minimum reliable \gls{trcd} values across different \gls{vpp} levels. Each curve represents a different DRAM module.}
    \label{fig:vpp_trcd}
\end{figure}

\noindent
\textbf{Experimental Results.}~\figref{fig:vpp_trcd} demonstrates {the variation in {\gls{trcdmin} (\secref{sec:experiment_design_trcd}) on the y-axis}}
{under} {reduced} \gls{vpp} {on the x-axis}{,} {across \nummodules{} DRAM modules}. 
We annotate the nominal \gls{trcd} value (\SI{13.5}{\nano\second})~\cite{jedec2017ddr4} with a black horizontal line. We make \obsvref{obsv:trcd_increase} from \figref{fig:vpp_trcd}.

\observation{{Reliable} row activation latency {generally} increases with reduced \gls{vpp}. {However, \numreliablechips{} {(\numreliablemodules{})} out of \numchips{} {(\nummodules{})} DRAM chips {(modules)}} complete {row} activation before the nominal activation latency.\label{obsv:trcd_increase}}

The minimum reliable activation latency {(\gls{trcdmin}) increases with reduced} \gls{vpp} across all tested modules. {\gls{trcdmin} exceeds the nominal \gls{trcd} of \SI{13.5}{\nano\second} for {\emph{only}} \numunreliablemodules{} of \nummodules{} tested modules (A0--A2, B2, and B5). Among these, modules from Mfr{.}~A and~B contain 16 and 8 chips per module. Therefore, we conclude that \numreliablechips{} of \numchips{} tested DRAM chips do \emph{not} experience bit flips when operated using nominal \gls{trcd}. We observe that since \gls{trcdmin} increases with reduced \gls{vpp}, the available \gls{trcd} guardband reduces by \trcdguardbandreduction{} with reduced \gls{vpp}{,} on average across all DRAM modules that reliably work with nominal \gls{trcd}. {{We also} observe that the three and two modules from Mfrs.~A and~B{, which exhibit \gls{trcdmin} values larger than the nominal \gls{trcd}}, 
reliably operate when {we use} a \gls{trcd} of \SI{24}{\nano\second} and \SI{15}{\nano\second}, respectively.}}

\begin{figure}[!ht]
    \centering
    \includegraphics[width=\linewidth]{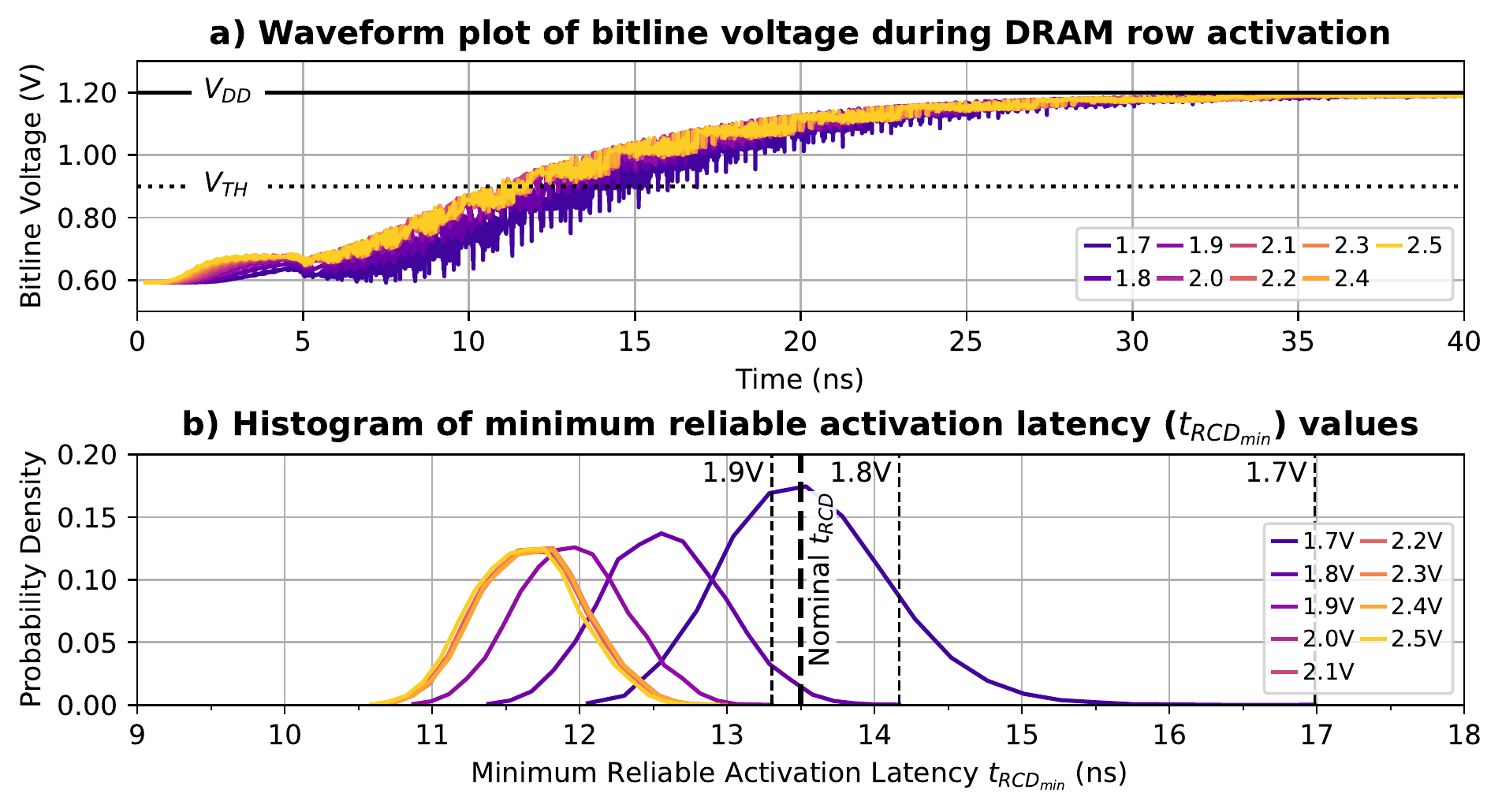}
    \caption{({a}) Waveform of the bitline voltage during row activation and ({b}) probability density distribution of {\gls{trcdmin}} values{,} {for} different \gls{vpp} levels.}
    \label{fig:spice_act_waveform}
\end{figure}

To verify our {experimental} observations and provide a deeper {insight} into the {effect of \gls{vpp} on} activation latency, we perform SPICE simulations {(}as described in \secref{sec:spice_model}{)}. \figref{fig:spice_act_waveform}a
shows a waveform of the bitline voltage during {the} row activation process. The time in the x-axis starts when {an} activation command is issued. Each color corresponds to the bitline voltage at a different \gls{vpp} level. We annotate the bitline's supply voltage (\gls{vdd}) and \gls{vthresh}. We make \obsvref{obsv:act_completes_later} from \figref{fig:spice_act_waveform}a.

\observation{{Row} activation successfully completes under reduced \gls{vpp} with an increased activation latency.\label{obsv:act_completes_later}}

{{\figref{fig:spice_act_waveform}a shows} that, as \gls{vpp} decreases, the bitline voltage takes longer to {increase} to \gls{vthresh}, resulting in a slower row activation. {{For example, \gls{trcdmin} increases from} \SI{11.6}{\nano\second} to \SI{13.6}{\nano\second} {(on average across $10^4$ Monte-Carlo simulation iterations)} {when \gls{vpp} is reduced from \SI{2.5}{\volt} to} \SI{1.7}{\volt}.}
{This happens {due to} two reasons. First, a lower \gls{vpp} creates a weaker channel in the access transistor, requiring a longer time for the capacitor and bitline to share charge. Second, the charge sharing process {(0--\SI{5}{\nano\second} in \figref{fig:spice_act_waveform}a)} {leads to} a smaller {change} in bitline voltage when \gls{vpp} is reduced due to the weakened charge restoration {process} that we explain in \secref{sec:sideeffects_charge_restoration}.}}

\figref{fig:spice_act_waveform}b shows the probability density distribution of \gls{trcdmin} values {under reduced \gls{vpp}} across {a total of $10^4$} Monte-Carlo simulation {iterations} {for different \gls{vpp} levels (color-coded)}. 
Vertical lines annotate the worst-case {reliable} \gls{trcdmin} values {across all iterations of {our} Monte-Carlo simulation (\secref{sec:spice_model})} for {different} \gls{vpp} levels.
We make \obsvref{obsv:trcd_spice_agrees} from \figref{fig:spice_act_waveform}b.

\observation{{SPICE simulations agree with our activation latency-related observations based on experiments on real DRAM chips{: \gls{trcdmin} increases with reduced \gls{vpp}.}\label{obsv:trcd_spice_agrees}}}

{We} analyze the variation in 
{1})~the probability density distribution of \gls{trcdmin}, and 
{2})~the worst-case (largest) {reliable} \gls{trcdmin} value when \gls{vpp} is reduced.
\figref{fig:spice_act_waveform}b shows that the probability density distribution of {\gls{trcdmin}} both shifts to larger values and {becomes} wider {with reduced \gls{vpp}}. 
{The} {worst-case (largest) \gls{trcdmin} increases from {\SI{12.9}{\nano\second} to \SI{13.3}{\nano\second}, \SI{14.2}{\nano\second}, and \SI{16.9}{\nano\second} when \gls{vpp} is reduced from \SI{2.5}{\volt} to \SI{1.9}{\volt}}, \SI{1.8}{\volt} and \SI{1.7}{\volt}, respectively.{\footnote{{SPICE simulation results do not show reliable operation when \gls{vpp}$\leq$\SI{1.6}{\volt}, yet real DRAM chips do operate reliably as we {show} in \secref{sec:sideeffects_trcd} and \secref{sec:sideeffects_retention}.}}} For a realistic nominal value of \SI{13.5}{\nano\second}, \gls{trcd}'s guardband reduces from {\SI{4.4}{\percent} to \SI{1.5}{\percent} as \gls{vpp} reduces from \SI{2.5}{\volt} to \SI{1.9}{\volt}}. {As \secref{sec:spice_model} explains, SPICE simulation results do \emph{not} exactly match {measured} real-device characteristics {(shown in \obsvref{obsv:trcd_increase})} because a SPICE model \emph{cannot} simulate a real DRAM chip's exact behavior without proprietary design and manufacturing information.}}

{From \obsvsref{obsv:trcd_increase}--\ref{obsv:trcd_spice_agrees}, we conclude that} 
1)~the {reliable row} activation latency increases with reduced \gls{vpp}, 
2)~the increase in {reliable row} activation latency does \emph{not} immediately require increasing {the} {nominal} \gls{trcd}{, but reduces the available guardband by \SI{21.9}{\percent}}
{for 208 out of 272 tested chips}{, and 3)~observed bit flips can be {eliminated} by increasing \gls{trcd} to \SI{24}{\nano\second} and \SI{15}{\nano\second} for erroneous modules from Mfrs.~A and~B.}

\subsection{DRAM Charge Restoration Under Reduced \gls{vpp}}
\label{sec:sideeffects_charge_restoration}

\head{\js{Motivation}}
A DRAM cell's charge restoration {process} is {affected by} \gls{vpp} because, similar to {the} row activation process, a DRAM cell capacitor's charge is restored through the channel formed in the access transistor, which is controlled by the wordline. Due to access transistor's characteristics, reducing \gls{vpp} without changing \gls{vdd} reduces \gls{vgs} and forms a weaker channel. To understand the impact of \gls{vpp} reduction on {the} charge restoration process, we investigate how charge restoration of a DRAM cell varies with {reduced} \gls{vpp}.

\begin{figure}[!ht]
    \centering
    \includegraphics[width=\linewidth]{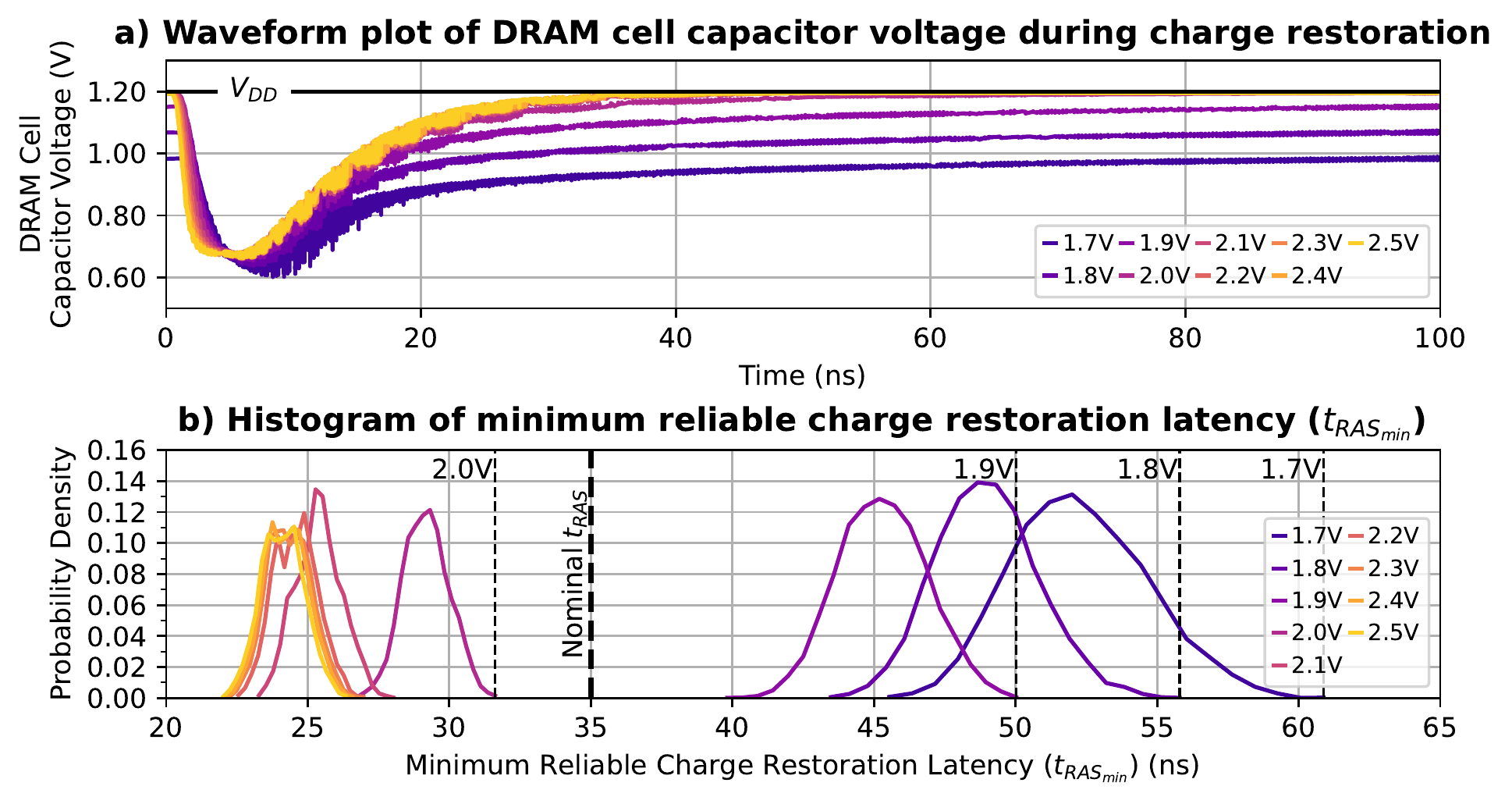}
    \caption{{(a)} Waveform of the {cell} capacitor voltage following a row activation and {(b)} probability density distribution of \agycrfour{\gls{trasmin}} values, {for} different \gls{vpp} levels.}
    \label{fig:charge_restoration_waveform}
\end{figure}

\head{{Experimental Results}}
Since our {FPGA} infrastructure {cannot probe} a DRAM cell capacitor's voltage level, we conduct this study in {our SPICE} simulation environment (\secref{sec:spice_model}).
\figref{fig:charge_restoration_waveform}a shows the waveform plot of capacitor voltage (y-axis) over time (x-axis), following a row activation event {(at t=0)}. \figref{fig:charge_restoration_waveform}b shows the probability density distribution (y-axis) of {\gls{trasmin} to {reliably} complete {the} charge restoration process on the x-axis} under different \gls{vpp} levels.
We make \obsvsref{obsv:restoration_chargereduction} and~\ref{obsv:restoration_largetras} from \figref{fig:charge_restoration_waveform}a and~\ref{fig:charge_restoration_waveform}b. 

\observation{A DRAM cell's capacitor voltage {can} saturate at a lower voltage level when \gls{vpp} is reduced.\label{obsv:restoration_chargereduction}}

{We observe that a DRAM cell capacitor's {voltage saturates at \gls{vdd} {(\SI{1.2}{\volt})} when \gls{vpp} is \SI{2.0}{\volt} or higher.}
However, the cell capacitor's voltage {saturates at a lower voltage level}
{by} \SI{4.1}{\percent}, \SI{11.0}{\percent}, and \SI{18.1}{\percent} {when \gls{vpp} is
\SI{1.9}{\volt}, \SI{1.8}{\volt}, and \SI{1.7}{\volt}, respectively.}}
This happens because the access transistor turns off when the voltage difference between its gate and source is smaller than a threshold {level}. For example, when \gls{vpp} is set to \SI{1.7}{\volt}, the access transistor allows charge restoration until the cell voltage reaches {\SI{0.98}{\volt}}. When the cell voltage reaches this level, the voltage difference between the gate (\SI{1.7}{\volt}) and the source {(\SI{0.98}{\volt})} is not large enough to form a strong channel, {causing} the cell voltage to saturate at {\SI{0.98}{\volt}}. This reduction in voltage can potentially {1)~increase the row activation latency (\gls{trcd}) and 2)~}reduce the cell's retention time. We {1)~already account for reduced saturation voltage's effect on \gls{trcd} in \secref{sec:sideeffects_trcd} and} 2)~investigate its effect on retention time in \secref{sec:sideeffects_retention}.

\observation{The increase in a DRAM cell's charge restoration latency with {reduced} \gls{vpp} {can} increase {the} \gls{tras} {timing parameter}, depending on the \gls{vpp} level.\label{obsv:restoration_largetras}}

{Similar to the variation in \gls{trcd} values that we discuss in \obsvref{obsv:trcd_spice_agrees}, the probability density distribution of \gls{tras} values also shifts to larger values {(i.e., \gls{tras} exceeds the nominal value when \gls{vpp} is lower than 2.0V)} and {becomes} wider {as} \gls{vpp} reduces. This happens as a {result} of reduced cell voltage, weakened channel in the access transistor, and reduced voltage level at the end of the charge sharing process, as we explain in \obsvref{obsv:trcd_spice_agrees}.}

{From \obsvsref{obsv:restoration_chargereduction} and~\ref{obsv:restoration_largetras}, we conclude that reducing \gls{vpp} {can} negatively affect the charge restoration process. {Reduced \gls{vpp}'s negative impact on charge restoration can potentially be mitigated {by} leveraging the guardbands in DRAM timing parameters~{\cite{lee2015adaptive,lee2017design, chang2016understanding, chang2017understanding, kim2018solar}} and using intelligent DRAM refresh techniques, where a partially restored DRAM row can be {refreshed} more frequently than other rows, {so that the row's charge is restored before it experiences a data retention bit flip}~\cite{das2018vrl, liu2012raidr, wang2018reducing}. We leave exploring such solutions {to} future work.}}

\subsection{DRAM Row Refresh Under Reduced \gls{vpp}}
\label{sec:sideeffects_retention}

\noindent
\textbf{Motivation.} \secref{sec:sideeffects_charge_restoration} demonstrates that the charge {re}stored in a DRAM cell {after a row activation} {can} be reduced as a result of \gls{vpp} reduction. This phenomenon is important for DRAM-based memories because reduced charge in {a} cell might reduce a DRAM cell's {data} retention time, {causing} \emph{retention bit flips} if the cell is \emph{not} refreshed more frequently. To understand the impact of \gls{vpp} reduction on real DRAM chips, we investigate the {effect of reduced \gls{vpp} on data retention related bit flips {using the methodology described in \secref{sec:experiment_design_retention}}.} 

\noindent
\textbf{Novelty.} This is the first work that experimentally analyzes the isolated impact of \gls{vpp} on DRAM cell retention {times}. Prior work~\cite{chang2017understanding} tests DDR3 DRAM chips under reduced \gls{vdd}, which may or may not change {the} internally{-}generated \gls{vpp} level.

\begin{figure}[!ht]
    \centering
    \includegraphics[width=\linewidth]{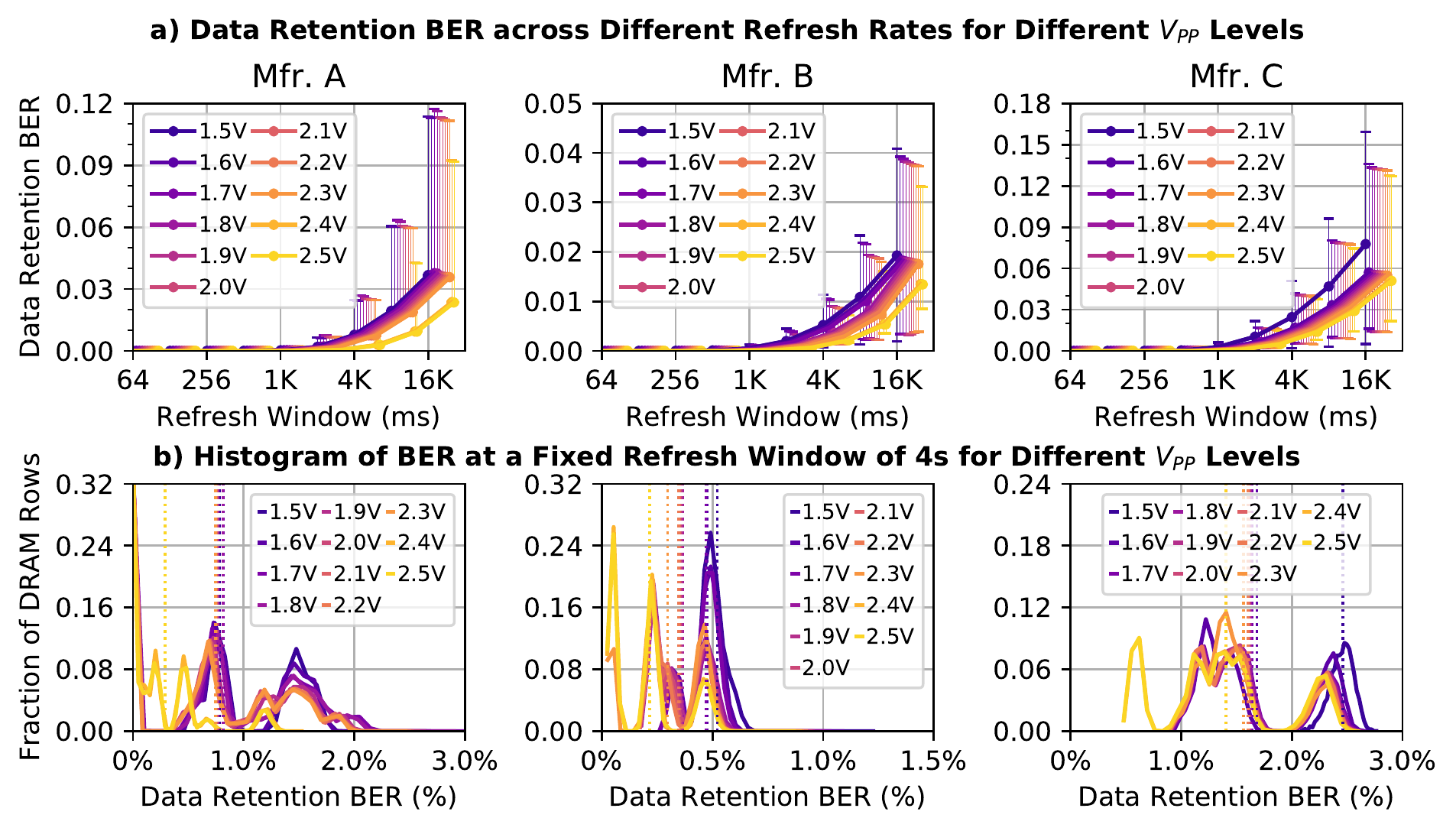}
    \caption{{Reduced \gls{vpp}'s effect on a) data retention \gls{ber} across different refresh rates and b) the distribution of data retention \gls{ber} across different DRAM rows for a fixed \gls{trefw} of \SI{4}{\second}.}}
    \label{fig:vpp_retention}
\end{figure}

\head{{Experimental Results}}
{\figref{fig:vpp_retention} demonstrates reduced \gls{vpp}'s effect on data retention \gls{ber} {on real DRAM chips}. \figref{fig:vpp_retention}a shows how the data retention \gls{ber} (y-axis) changes with increasing refresh window (log-scaled in x-axis) for different \gls{vpp} levels (color-coded). Each curve in \figref{fig:vpp_retention}a shows the average \gls{ber} across all DRAM rows, and error bars mark {the} \SI{90}{\percent} confidence interval. {The x-axis starts from \SI{64}{\milli\second} because we do \emph{not} observe any bit flips at \gls{trefw} values {smaller} than \SI{64}{\milli\second}}. To provide deeper insight into reduced \gls{vpp}'s effect on data retention \gls{ber}, \figref{fig:vpp_retention}b demonstrates the population density distribution of data retention \gls{ber} across tested rows for {a} \gls{trefw} of \SI{4}{\second}. Dotted vertical lines mark the average \gls{ber} across rows for each \gls{vpp} level.}
We make \obsvsref{obsv:retention_vpp} and~\ref{obsv:retention_minreliable} from \figref{fig:vpp_retention}.

\observation{More DRAM cells {tend to} experience {data} retention bit flips when \gls{vpp} is reduced.\label{obsv:retention_vpp}}

{\figref{fig:vpp_retention}a shows that data retention \gls{ber} curve is higher (e.g., dark-purple compared to yellow) for smaller \gls{vpp} levels (e.g., \SI{1.5}{\volt} compared to \SI{2.5}{\volt}). 
{To provide a deeper insight, \figref{fig:vpp_retention}b shows that} average data retention \gls{ber} {across all tested rows when \gls{trefw}=\SI{4}{\second}} increases from \SI{0.3}{\percent}, \SI{0.2}{\percent}, and \SI{1.4}{\percent} {for a} \gls{vpp} {of} \SI{2.5}{\volt} to \SI{0.8}{\percent}, \SI{0.5}{\percent}, and \SI{2.5}{\percent} {for a} \gls{vpp} of \SI{1.5}{\volt} for Mfrs.~A, B, and~C, respectively.}
We {hypothesize} that this happens because of the weakened charge restoration process {with reduced \gls{vpp} (\secref{sec:sideeffects_charge_restoration})}.

\observation{{Even though} DRAM cells experience retention bit flips at smaller retention times when \gls{vpp} is reduced, {23} of {30} {tested} modules {experience \emph{no} data} retention bit flips {at} the nominal refresh window (\SI{64}{\milli\second}).\label{obsv:retention_minreliable}}

{Data retention \gls{ber} is {very} low at the \gls{trefw} of \SI{64}{\milli\second} even for {a} \gls{vpp} of \SI{1.5}{\volt}. We observe that \emph{no} DRAM module from Mfr.~A exhibits a data retention bit flip at {the} \SI{64}{\ms} \gls{trefw}, and \emph{only} three and four modules from Mfrs.~B (B6, B8, and B9) and~C (C1, C3, C5, and C9) experience bit flips across all 30 DRAM modules we test.}

{We} investigate the significance of the{ {observed} data} retention bit flips {and whether it is possible to mitigate these bit flips using error correcting codes (ECC)~\cite{hamming1950error} or {other existing methods to avoid data retention bit flips (e.g., selectively refreshing a small fraction of DRAM rows at a higher refresh rate}~\cite{das2018vrl, liu2012raidr, wang2018reducing}).}
{To do so, we analyze the nature of data retention bit flips {when each tested module is operated at the module's \gls{vppmin}}
{for two \gls{trefw} values: \SI{64}{\milli\second} and \SI{128}{\milli\second}, which are the smallest refresh windows that yield non-zero BER for different DRAM modules.} 

{To evaluate whether data retention bit flips can be avoided using ECC, we assume a realistic data word size of 64 bits~\cite{cojocar2019eccploit, kim2016all, patel2019understanding, patel2020beer, patel2021harp, patel2022case, mineshphd}. We make \obsvref{obsv:retention_onlyone} from this analysis.}}

\observation{{Data retention errors can be avoided using simple single error correcting codes} at the smallest \gls{trefw} that yield{s} non-zero \gls{ber}.\label{obsv:retention_onlyone}}

{We observe that \emph{no} 64-bit data word contains more than one bit flip for the smallest \gls{trefw} that yield non-zero \gls{ber}.} 
{We conclude that simple \emph{single error correction double error detection (SECDED) ECC} can correct \emph{all} erroneous data words.}

\begin{figure}[!ht]
    \centering
    \includegraphics[width=\linewidth]{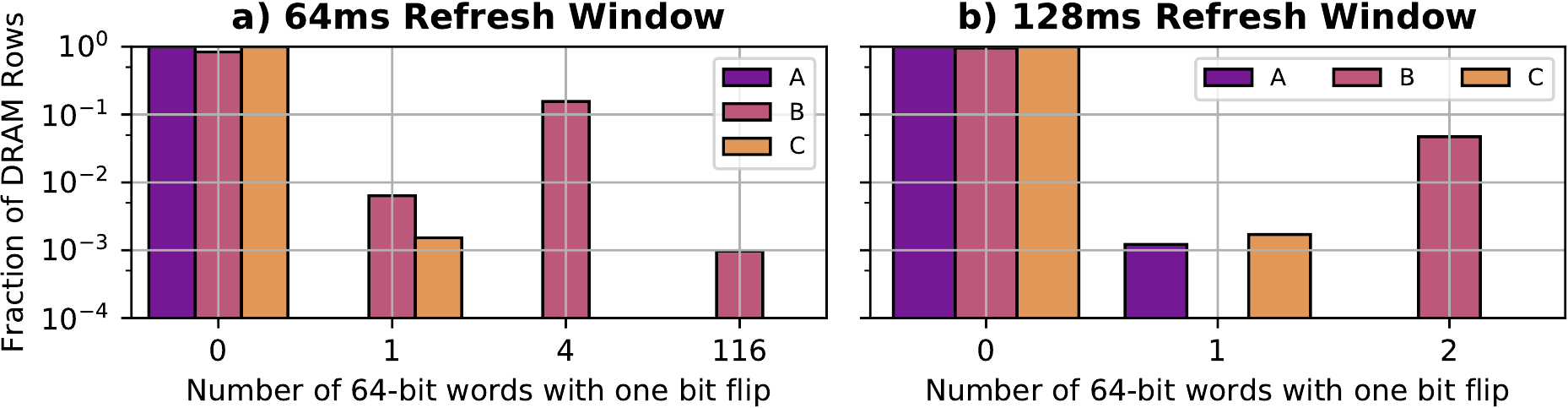}
    \caption{{Data retention bit flip characteristics of DRAM rows in DRAM modules that exhibit bit flips at (a) \SI{64}{\milli\second} and (b) \SI{128}{\milli\second} refresh windows but not at lower \gls{trefw} values when operated at \gls{vppmin}. Each subplot shows the distribution of DRAM rows based on the number of erroneous {64-bit} words that the rows exhibit.}}
    \label{fig:vpp_retention_ber_hist}
\end{figure}

{To evaluate whether data retention bit flips can be avoided {by} selectively refreshing a small fraction of DRAM rows, we analyze the distribution of these bit flips across different DRAM rows. \figref{fig:vpp_retention_ber_hist}a (\figref{fig:vpp_retention_ber_hist}b) shows the distribution of DRAM rows {that} experience a data retention bit flip when \gls{trefw} is} \SI{64}{\milli\second} 
(\SI{128}{\milli\second})
{but \emph{not} at a smaller \gls{trefw}, based on their data retention bit flip characteristics}.
{The x-axis shows the number of 64-bit data words with one bit flip in a DRAM row.}
{The y-axis shows the fraction of DRAM rows in log-scale, exhibiting the behavior, specified in the x-axis {for different manufacturers (color-coded)}.
} 
We make \obsvref{obsv:retention_smallfraction} {from \figref{fig:vpp_retention_ber_hist}.}

\observation{{Only a small fraction (\SI{16.4}{\percent} / \SI{5.0}{\percent}) of DRAM rows  contain erroneous data words at the smallest \gls{trefw} (\SI{64}{\milli\second} / \SI{128}{\milli\second}) that yields non-zero \gls{ber}.}\label{obsv:retention_smallfraction}}

{\figref{fig:vpp_retention_ber_hist}a} shows that modules from Mfr.~A do \emph{not} exhibit any bit flips when \gls{trefw} is \SI{64}{\milli\second}, while \SI{15.5}{\percent} and \SI{0.2}{\percent} of DRAM rows in modules from Mfrs.~B and~C exhibit four and one 64-{bit} words with a single bit flip, respectively{; and \SI{0.01}{\percent} of DRAM rows from Mfr.~B contain 116 data words with one bit flip.} {\figref{fig:vpp_retention_ber_hist}b} shows that \SI{0.1}{\percent}, \SI{4.7}{\percent}, and \SI{0.2}{\percent} of rows from Mfrs.~A, B, and~C contain 1, 2, and 1 erroneous {data words, respectively,}
when the refresh window is \SI{128}{\milli\second}. {We conclude that \emph{all} of these data retention bit flips can be avoided}
{by doubling the refresh rate\footnote{{We test our chips {at} fixed refresh rates in increasing powers of two (\secref{sec:experiment_design_retention}). {Therefore, our experiments do \emph{not} capture whether eliminating a bit flip is possible by increasing the}
refresh rate by less than $2\times$. We leave a finer granularity data retention time analysis to future work.}} \emph{only} for} {\SI{16.4}{\percent} / \SI{5.0}{\percent} of DRAM} rows~\cite{das2018vrl, liu2012raidr, wang2018reducing} when \gls{trefw} is \SI{64}{\milli\second} / \SI{128}{\milli\second}.

From \obsvsref{obsv:retention_vpp}--\ref{obsv:retention_smallfraction}, we conclude that a DRAM row's {data} retention time can reduce when \gls{vpp} is reduced. However,
{1)~{most of (i.e.,} 23 out of 30{)} tested modules do \emph{not} exhibit any bit flips {at} the nominal \gls{trefw} of \SI{64}{\milli\second} and 2)~bit flips observed in seven modules can be mitigated using {existing {SECDED} ECC~\cite{hamming1950error} or selective refresh methods~\cite{das2018vrl, liu2012raidr, wang2018reducing}}.}

\section{{Limitations of Wordline Voltage Scaling}}
\label{sec:limitations}

{We highlight \param{{four}} key limitations of wordline voltage scaling and our experimental characterization. 

First, in our experiments, we observe that none of the tested DRAM modules reliably operate at a \gls{vpp} lower than {a certain {voltage level, called \gls{vppmin}}}. This happens because an access transistor cannot connect the DRAM cell capacitor to the bitline when the access transistor’s gate-to-source voltage difference is \emph{not} larger than the transistor’s threshold voltage. Therefore, each DRAM chip has a minimum \gls{vpp} level {at} which it can reliably operate {(e.g., {lowest at} \SI{1.4}{\volt} for A0 and {highest at} {\SI{2.4}{\volt} for A5}).} 
With this limitation,} we observe \hcfirstincravg{} {/} \berdecravg{} {average} {increase /} reduction in {\gls{hcfirst} /} \gls{ber} {across all tested DRAM chips at {their respective} \gls{vppmin} {levels}. A DRAM chip's RowHammer vulnerability can potentially reduce {further} if access transistors are designed to {operate at} smaller \gls{vpp} levels.} 

Second, 
we cannot investigate the root cause of all results {we observe} since {1)~}DRAM manufacturers do \emph{not} {describe} the exact circuit design {details of} their commodity DRAM chips~\cite{mineshphd,saroiu2022price,frigo2020trrespass,patel2022case} {{and} 2)~our infrastructure{'s} physical limitations {{prevent} us from} observing a DRAM chip's {exact internal} behavior (e.g., it is \emph{not} possible to directly measure a cell's capacitor voltage).} 

{Third, this paper does \emph{not} thoroughly analyze the {three-way} interaction between \gls{vpp}, temperature, and RowHammer. {T}here is {already} a complex {two-way} interaction between RowHammer and temperature, requiring {studies} to test each DRAM cell at all {allowed} temperature levels~\cite{orosa2021deeper}. Since {a three-way interaction study requires even more} characterization {that would take} several months {of} testing time, we leave it to future work to study the interaction between \gls{vpp}, temperature, and RowHammer.}

{Fourth}, we experimentally demonstrate that the RowHammer vulnerability can be mitigated by {reducing \gls{vpp} at the cost of {a} \trcdguardbandreduction{} {average} reduction in the \gls{trcd} guardband of tested DRAM chips. 
{Although reducing the guardband can hurt DRAM manufacturing yield, we leave studying \gls{vpp} reduction's effect on yield to future work because we do \emph{not} have access to DRAM manufacturers' proprietary yield statistics.}}

\section{{Key Takeaways}}
\label{sec:takeaways}

{We} summarize the key findings of our {experimental} analyses of the {\acrlong{vpp} (\gls{vpp})'s effect on the RowHammer vulnerability and reliabl{e operation} of modern DRAM chips.} 
{From our new observations, we draw two key takeaways.}

\noindent
\textbf{Takeaway 1: {Effect} of \gls{vpp} on RowHammer.} We observe that {scaling down} \gls{vpp} reduces a DRAM chip's RowHammer vulnerability, such that {RowHammer} \gls{ber} {\emph{decreases} by \berdecravg{} (up to \berdecrmax{}) and \gls{hcfirst} increases by \hcfirstincravg{} (up to \hcfirstincrmax{}) on average across all DRAM rows}{. Only} \fracberopposingrows{} and \frachcfirstopposingrows{} of DRAM rows exhibit 
opposite {\gls{ber} and \gls{hcfirst} trends, respectively} {(\secref{sec:vpp:ber} and \secref{sec:vpp_vs_hcfirst})}.

\noindent
{\textbf{Takeaway 2: {Effect of \gls{vpp} on} DRAM {reliability}.} We observe that reducing \gls{vpp}
{1)~reduces the existing guardband for row activation {latency} by \trcdguardbandreduction{} on average across tested chips}
and 2)~causes DRAM cell charge to saturate at {\SI{1}{\volt}} instead of \SI{1.2}{\volt} (\gls{vdd}) (\secref{sec:sideeffects_charge_restoration}),
leading \SI{0}{\percent}, \SI{15.5}{\percent}, and \SI{0.2}{\percent} of DRAM rows to experience {SECDED ECC-correctable} data retention {bit flips} at {the} nominal refresh window of \SI{64}{\milli\second} {in DRAM modules from Mfrs.~A, B, and~C, respectively} (\secref{sec:sideeffects_retention}).}

{{\head{{Finding} Optimal Wordline Voltage} Our {two key takeaways suggest that reducing RowHammer vulnerability of a DRAM chip {via} \gls{vpp} reduction}
can require {1)~accessing DRAM rows with a slightly larger latency, 2)~employing error correcting codes (ECC), or 3)~refreshing a small subset of rows at a higher refresh rate.}
Therefore, one can define different Pareto-optimal operating conditions for different performance and reliability requirements. For example, a security-critical system can choose a lower \gls{vpp} to reduce RowHammer vulnerability, whereas a performance-critical and error-{tolerant} system {might} prefer lower access latency over {higher RowHammer tolerance}.}} {DRAM designs and systems that are informed about the tradeoffs between \gls{vpp}, access latency, and retention time can make better{-}informed design decisions (e.g., fundamentally enable lower access latency) or employ better{-}informed memory controller policies (e.g., {using longer \gls{trcd},} {employing SECDED ECC, or} {doubling the refresh rate only for a small fraction of rows when the chip operates at} reduced \gls{vpp}).} {We believe such {designs} are important to explore in future work.} We hope that {the new insights we provide} can lead to the design of {stronger DRAM-based systems against RowHammer {along with {better-}informed DRAM{-based system} designs}.}
 
\section{Related Work}
\label{sec:related_work}
{{To our knowledge, this} is the first work that 
{experimentally studies how reducing {wordline voltage} affects a real DRAM chip's 1)~}RowHammer vulnerability{, 2)~row activation latency, 3)~charge restoration process, and 4)~data retention time.}~{We divide prior work into {three} categories: 1)~{explorations} of reduced-voltage DRAM operation, 2)~experimental characterization studies {of the} RowHammer vulnerability of real DRAM chips, {{and}} 3)~RowHammer attacks and defenses.}
}

\noindent
{\textbf{Reduced-Voltage DRAM {Operation}.}~Prior works~\cite{david2011memory, deng2011memscale, chang2017understanding} propose operating DRAM {with} reduced \gls{vdd} to improve energy efficiency. {\cite{david2011memory} and \cite{deng2011memscale} propose dynamic voltage and frequency scaling (DVFS) {for DRAM chips} and \cite{david2011memory} provides results in a real system.}
{\cite{chang2017understanding} proposes to scale down \gls{vdd} without reducing DRAM chip frequency. To do so, \cite{chang2017understanding} experimentally demonstrates the interaction between \gls{vdd} and DRAM row access latency in real DDR3 DRAM chips.}
{These three works neither focus on the RowHammer vulnerability nor distinguish {between} \gls{vdd} and \gls{vpp}.} {Unlike {these works}, w}e focus on the impact of \gls{vpp} (isolated from \gls{vdd}) on RowHammer {and reliable operation} characteristics of real DDR4 DRAM chips.}

\noindent
{\textbf{{Experimental} RowHammer Characterization.} Prior work{s} extensively characterize the RowHammer vulnerability in real DRAM chips\cite{orosa2021deeper, hassan2021utrr,kim2020revisiting, kim2014flipping, park2016experiments, frigo2020trrespass}. These works {experimentally demonstrate {(}using real DDR3, DDR4, and LPDDR4 DRAM chips how{)} a DRAM chip's RowHammer vulnerability varies with}
1)~DRAM refresh rate~\cite{hassan2021utrr,frigo2020trrespass,kim2014flipping}, 2)~the physical distance between aggressor and victim rows~\cite{kim2014flipping,kim2020revisiting}, 3)~DRAM generation and technology node~\cite{orosa2021deeper,kim2014flipping,kim2020revisiting,hassan2021utrr}, 4)~temperature~\cite{orosa2021deeper,park2016experiments}, 5)~the time the aggressor row stays active~\cite{orosa2021deeper,park2016experiments}, and~6)~physical location of the victim DRAM cell~\cite{orosa2021deeper}. {N}one of these works analyze how reduced \gls{vpp} {affects} RowHammer vulnerability in real DRAM chips. Our characterization study furthers the analyses in these works by uncovering new insights into RowHammer behavior {and DRAM operation}.}

\noindent
{\textbf{RowHammer Attacks and Defenses.} 
{Many} prior works~\cite{seaborn2015exploiting, van2016drammer, gruss2016rowhammer, razavi2016flip, pessl2016drama, xiao2016one, bosman2016dedup, bhattacharya2016curious, qiao2016new, jang2017sgx, aga2017good,
mutlu2017rowhammer, tatar2018defeating ,gruss2018another, lipp2018nethammer,
van2018guardion, frigo2018grand, cojocar2019eccploit,  ji2019pinpoint, mutlu2019rowhammer, hong2019terminal, kwong2020rambleed, frigo2020trrespass, cojocar2020rowhammer, weissman2020jackhammer, zhang2020pthammer, rowhammergithub, yao2020deephammer, deridder2021smash, hassan2021utrr, jattke2022blacksmith, marazzi2022protrr, tol2022toward,burleson2016invited, brasser2017can,qureshi2021rethinking, orosa2021deeper, redeker2002investigation, saroiu2022price, yang2019trap, walker2021ondramrowhammer, park2016statistical, kim2020revisiting, kim2014flipping, park2016experiments} show that RowHammer can be exploited to mount system-level attacks to compromise system security {and safety} (e.g., to acquire root privileges {or leak private data}).
To protect against these attacks,
{many prior works~\cite{aichinger2015ddr, AppleRefInc, aweke2016anvil, kim2014flipping, kim2014architectural,son2017making, lee2019twice, you2019mrloc, seyedzadeh2018cbt, van2018guardion, konoth2018zebram, park2020graphene, yaglikci2021blockhammer, kang2020cattwo, bains2015row, bains2016distributed, bains2016row, brasser2017can, gomez2016dummy, jedec2017ddr4,hassan2019crow, devaux2021method, ryu2017overcoming, yang2016suppression, yang2017scanning, gautam2019row, yaglikci2021security, qureshi2021rethinking, greenfield2012throttling, marazzi2022protrr, saileshwar2022randomized} propose RowHammer mitigation {mechanisms} that prevent RowHammer bit flips from compromising a system.}
The novel observations {we make in this work} can be leveraged to reduce RowHammer vulnerability {and} complement existing RowHammer defense mechanisms{,} further increas{ing} their effectiveness and reduc{ing} their overheads.
}

\glsresetall{}
\section{Conclusion}
\label{sec:conclusion}

{{W}e present the first experimental RowHammer characterization study under reduced {\gls{vpp}}. Our results, using \numchips{} real DDR4 DRAM chips from three major manufacturers, show that {RowHammer vulnerability can be reduced by reducing} \gls{vpp}. 
Using real-device experiments and SPICE simulations, we demonstrate that although the reduced \gls{vpp} slightly worsens DRAM access latency, charge restoration process and data retention time, {most of (}\numreliablechips{} out of \numchips{}{) tested chips} reliably work under reduced \gls{vpp} {leveraging} already existing guardbands of nominal timing parameters {and employing existing ECC or selective refresh techniques}.} Our findings provide {new} insights into the increasingly {critical} RowHammer problem in modern DRAM {chips. We hope that they} lead to the design of {systems that are more robust against RowHammer attacks.}
\section*{Acknowledgments}
We thank our shepherd Karthik Pattabiraman and the anonymous reviewers of DSN~2022 for valuable feedback. We thank the SAFARI Research Group members for valuable feedback and the stimulating intellectual environment they provide. We acknowledge the generous gifts provided by our industrial partners{, including} Google, Huawei, Intel, Microsoft, and VMware{, and support from {the} Microsoft Swiss Joint Research Center}.

\balance

\bibliographystyle{IEEEtran}
\bibliography{refs}
\pagebreak
\nobalance
\onecolumn
\glsresetall{}
\setstretch{1.0}
\appendix
\renewcommand{\thesection}{Appendix \Alph{section}}
\section{\agyextthree{T}ested DRAM Modules}
\label{sec:appendix}

\agyextone{Table~\ref{tab:detailed_dimm_table} shows the characteristics of the \omexttwo{DDR4} DRAM modules we test and analyze.\footnote{\hluoextone{\agyexttwo{All tested DRAM modules implement the DDR4 DRAM standard~\cite{jedec2017ddr4}.} We make our best effort in identifying the DRAM chips used in our tests. We identify the DRAM chip density and die revision through the original manufacturer markings on the chip. For certain DIMMs we tested, the original DRAM chip markings are removed by the DIMM manufacturer. In this case, we can only identify the chip manufacturer and density by reading the information stored in the SPD. However, these DIMM manufacturers also tend to remove the die revision information in the SPD. Therefore, we \emph{cannot} identify the die revision \agyexttwo{of five DIMMs} and \agyexttwo{the} manufacturing date \agyexttwo{of} six DIMMs we test\agyexttwo{, shown as `-' in the table}.}} For each DRAM module, we provide the 1)~DRAM chip manufacturer, 2)~\agyexttwo{DIMM name, 3)~DIMM model,\footnote{\hluoextone{DIMM models CMV4GX4M1A2133C1\agyextthree{5} and F4-2400C17S-8GNT appear in more than one DRAM chip manufacturer because different batches of these modules use DRAM chips from different manufacturers (i.e., Micron-SK Hynix and Samsung-SK Hynix, respectively) across different batches.}} 4)}~die density, \agyexttwo{5)~data transfer frequency, 6)~chip organization,} \agyexttwo{7})~die revision, specified in the module's serial presence detect (SPD) registers, \agyexttwo{8)~manufacturing} date, specified on the module's label \agyexttwo{in the form of $week-year$, and 9)~RowHammer vulnerability characteristics of the module}. 
\agyexttwo{Table~\ref{tab:detailed_dimm_table} reports the RowHammer vulnerability characteristics of each DIMM under two \gls{vpp} levels: \emph{i)}~nominal \gls{vpp} (\SI{2.5}{\volt}) and \emph{ii)~}\gls{vppmin}. We quantify a DIMM's RowHammer vulnerability characteristics at a given \gls{vpp} in terms of two metrics: \emph{i)}~\gls{hcfirst} and \emph{ii)}~\gls{ber}. Based on these two metrics at nominal \gls{vpp} and \gls{vppmin}, Table~\ref{tab:detailed_dimm_table} \agyextthree{also} provides a \emph{recommended} \gls{vpp} level ($V_{PP_{Rec}}$) and the corresponding RowHammer characteristics in the right-most three columns.}
}
\vspace{1.5em}

\newcommand{\rotcell}[1]{\rotatebox{90}{#1}}

\begin{table}[h!]
\centering
\caption{Tested DRAM modules and their characteristics when \gls{vpp}=\SI{2.5}{\volt} (nominal) and \gls{vpp}=\gls{vppmin}. \gls{vppmin} is specified for each module.}
\resizebox{\columnwidth}{!}{
\begin{tabular}{|l|c|l|ccccc||rc|c|rc|c|rc|} 
\hline
& & & & & & & & \multicolumn{2}{c|}{\textbf{$\mathbf{V_{PP}}$ = 2.5V}} & & \multicolumn{2}{c|}{$\mathbf{V_{PP} = V_{PP_{min}}}$} & & \multicolumn{2}{c|}{$\mathbf{V_{PP} = V_{PP_{Rec}}}$}\\
\rotcell{\begin{tabular}[c]{@{}c@{}}\textbf{DRAM Chip Mfr.}\end{tabular}} & \rotcell{\begin{tabular}[c]{@{}c@{}}\textbf{DIMM Name}\end{tabular}} & \textbf{DIMM Model} & \rotcell{\begin{tabular}[c]{@{}c@{}}\textbf{Die Density}\end{tabular}} & \rotcell{\begin{tabular}[c]{@{}c@{}}\textbf{Freq\agyextthree{uency} (MT/s)}\end{tabular}} & \rotcell{\textbf{Chip Org.}} & \rotcell{\begin{tabular}[c]{@{}c@{}}\textbf{Die Revision}\end{tabular}} & \rotcell{\begin{tabular}[c]{@{}c@{}}\textbf{Mfr. Date}\end{tabular}} &\rotcell{\begin{tabular}[c]{@{}l@{}}\textbf{Minimum}\\\textbf{$\mathbf{HC_{first}}$}\end{tabular}}& \textbf{$\mathbf{BER}$}& \rotcell{$\mathbf{V_{PP_{min}}}$} &\rotcell{\begin{tabular}[c]{@{}l@{}}\textbf{Minimum}\\\textbf{$\mathbf{HC_{first}}$}\end{tabular}}&\textbf{$\mathbf{BER}$}& \rotcell{\begin{tabular}[c]{@{}l@{}}\textbf{Recommended}\\\textbf{$\mathbf{V_{PP} (V_{PP_{Rec}}}$)}\end{tabular}} & \rotcell{\begin{tabular}[c]{@{}l@{}}\textbf{Minimum}\\\textbf{$\mathbf{HC_{first}}$}\end{tabular}}&\textbf{$\mathbf{BER}$}\\ 
\hline
\hline
\multirow{10}{*}{\rotcell{\begin{tabular}[c]{@{}l@{}}Mfr. A (Micron)\end{tabular}}}                                                            & A0                                                                                               & MTA18ASF2G72PZ-2G3B1QK~\cite{datasheetMTA18ASF2G72PZ}               & 8Gb                                                                                                & 2400                                             & x4                                                    & B                                                                                               & 11-19                                                                                            & 39.8K                                                                                         & 1.24e-03                                             & 1.4                                                                                                                                                             & 42.2K                                                                                         & 1.00e-03                                             & 1.4                                                                                                                                                 & 42.2K                                   & 1.00e-03                  \\
                                                                                                                                     & A1                                                                                               & MTA18ASF2G72PZ-2G3B1QK~\cite{datasheetMTA18ASF2G72PZ}               & 8Gb                                                                                                & 2400                                             & x4                                                    & B                                                                                               & 11-19                                                                                            & 42.2K                                                                                         & 9.90e-04                                             & 1.4                                                                                                                                                             & 46.4K                                                                                         & 7.83e-04                                             & 1.4                                                                                                                                                 & 46.4K                                   & 7.83e-04                  \\
                                                                                                                                     & A2                                                                                               & MTA18ASF2G72PZ-2G3B1QK~\cite{datasheetMTA18ASF2G72PZ}               & 8Gb                                                                                                & 2400                                             & x4                                                    & B                                                                                               & 11-19                                                                                            & 41.0K                                                                                         & 1.24e-03                                             & 1.7                                                                                                                                                             & 39.8K                                                                                         & 1.35e-03                                             & 2.\agyextthree{1}                                                                                                                                                 & 4\agyextthree{2.1}K                                   & 1.\agyextthree{55e-3}                  \\
                                                                                                                                     & A3                                                                                               & CT4G4DFS8266.C8FF~\cite{datasheetCT4G4DFS8266}                    & 4Gb                                                                                                & 2666                                             & x8                                                    & F                                                                                               & 07-21                                                                                            & 16.7K                                                                                         & 3.33e-02                                             & 1.4                                                                                                                                                             & 16.5K                                                                                         & 3.52e-02                                             & \agyextthree{1.7}                                                                                                                                                 & \agyextthree{17.0K}                                   & 3.\agyextthree{48}e-02                  \\
                                                                                                                                     & A4                                                                                               & CT4G4DFS8266.C8FF~\cite{datasheetCT4G4DFS8266}                    & 4Gb                                                                                                & 2666                                             & x8                                                    & F                                                                                               & 07-21                                                                                            & 14.4K                                                                                         & 3.18e-02                                             & 1.5                                                                                                                                                             & 14.4K                                                                                         & 3.33e-02                                             & 2.5                                                                                                                                                 & 14.4K                                   & 3.18e-02                  \\
                                                                                                                                     & A5                                                                                               & CT4G4SFS8213.C8FBD1                  & 4Gb                                                                                                & 2400                                             & x8                                                    & -                                                                                               & 48-16                                                                                            & 140.7K                                                                                        & 1.39e-06                                             & 2.4                                                                                                                                                             & 145.4K                                                                                        & 3.39e-06                                             & 2.4                                                                                                                                                 & 145.4K                                  & 3.39e-06                  \\
                                                                                                                                     & A6                                                                                               & CT4G4DFS8266.C8FF~\cite{datasheetCT4G4DFS8266}                    & 4Gb                                                                                                & 2666                                             & x8                                                    & F                                                                                               & 07-21                                                                                            & 16.5K                                                                                         & 3.50e-02                                             & 1.5                                                                                                                                                             & 16.5K                                                                                         & 3.66e-02                                             & 2.5                                                                                                                                                 & 16.5K                                   & 3.50e-02                  \\
                                                                                                                                     & A7                                                                                               & CMV4GX4M1A2133C15~\cite{datasheetCMV4GX4M1A2133C15}                     & 4Gb                                                                                                & 2133                                             & x8                                                    & -                                                                                               & -                                                                                                & 16.5K                                                                                         & 3.42e-02                                             & 1.8                                                                                                                                                             & 16.5K                                                                                         & 3.52e-02                                             & 2.5                                                                                                                                                 & 16.5K                                   & 3.42e-02                  \\
                                                                                                                                     & A8                                                                                               & MTA18ASF2G72PZ-2G3B1QG~\cite{datasheetMTA18ASF2G72PZ}               & 8Gb                                                                                                & 2400                                             & x4                                                    & B                                                                                               & 11-19                                                                                            & 35.2K                                                                                         & 2.38e-03                                             & 1.4                                                                                                                                                             & 39.8K                                                                                         & 2.07e-03                                             & 1.4                                                                                                                                                 & 39.8K                                   & 2.07e-03                  \\
                                                                                                                                     & A9                                                                                               & CMV4GX4M1A2133C15~\cite{datasheetCMV4GX4M1A2133C15}                    & 4Gb                                                                                                & 2133                                             & x8                                                    & -                                                                                               & -                                                                                                & 14.3K                                                                                         & 3.33e-02                                             & 1.5                                                                                                                                                             & 14.3K                                                                                         & 3.48e-02                                             & \agyextthree{1.6}                                                                                                                                                 & 14.\agyextthree{6}K                                   & 3.\agyextthree{47}e-02                  \\ 
\hline
\multirow{10}{*}{\rotcell{\begin{tabular}[c]{@{}l@{}}Mfr. B (Samsung)\end{tabular}}}                                                           & B0                                                                                               & M378A1K43DB2-CTD~\cite{datasheetM378A1K43DB2}                     & 8Gb                                                                                                & 2666                                             & x8                                                    & D                                                                                               & 10-21                                                                                            & 7.9K                                                                                          & 1.18e-01                                             & 2.0                                                                                                                                                             & 7.6K                                                                                          & 1.22e-01                                             & 2.5                                                                                                                                                 & 7.9K                                    & 1.18e-01                  \\
                                                                                                                                     & B1                                                                                               & M378A1K43DB2-CTD~\cite{datasheetM378A1K43DB2}                     & 8Gb                                                                                                & 2666                                             & x8                                                    & D                                                                                               & 10-21                                                                                            & 7.3K                                                                                          & 1.26e-01                                             & 2.0                                                                                                                                                             & 7.6K                                                                                          & 1.28e-01                                             & 2.0                                                                                                                                                 & 7.6K                                    & 1.28e-01                  \\
                                                                                                                                     & B2                                                                                               & F4-2400C17S-8GNT~\cite{datasheetF42400C17S8GNT}                     & 4Gb                                                                                                & {2400}         & x8                                                    & F                                                                                               & 02-21                                                                                            & 11.2K                                                                                         & 2.52e-02                                             & 1.6                                                                                                                                                             & 12.0K                                                                                         & 2.22e-02                                             & 1.6                                                                                                                                                 & 12.0K                                   & 2.22e-02                  \\
                                                                                                                                     & B3                                                                                               & M393A1K43BB1-CTD6Y~\cite{datasheetM393A1K43BB1}                   & 8Gb                                                                                                & 2666                                             & x8                                                    & B                                                                                               & 52-20                                                                                            & 16.6K                                                                                         & 2.73e-03                                             & 1.6                                                                                                                                                             & 21.1K                                                                                         & 1.09e-03                                             & 1.6                                                                                                                                                 & 21.1K                                   & 1.09e-03                  \\
                                                                                                                                     & B4                                                                                               & M393A1K43BB1-CTD6Y~\cite{datasheetM393A1K43BB1}                   & 8Gb                                                                                                & 2666                                             & x8                                                    & B                                                                                               & 52-20                                                                                            & 21.0K                                                                                         & 2.95e-03                                             & 1.8                                                                                                                                                             & 19.9K                                                                                         & 2.52e-03                                             & 2.\agyextthree{0}                                                                                                                                                 & 21.\agyextthree{1}K                                   & 2.\agyextthree{68}e-03                  \\
                                                                                                                                     & B5                                                                                               & M471A5143EB0-CPB~\cite{datasheetM471A5143EB0}                     & 4Gb                                                                                                & 2133                                             & x8                                                    & E                                                                                               & 08-17                                                                                            & 21.0K                                                                                         & 7.78e-03                                             & 1.8                                                                                                                                                             & 21.0K                                                                                         & 6.02e-03                                             & \agyextthree{2.0}                                                                                                                                                 & 21.\agyextthree{1}K                                   & \agyextthree{8.67}e-03               \\
                                                                                                                                     & B6                                                                                               & CMK16GX4M2B3200C16~\cite{datasheetCMK16GX4M2B3200C16}                   & 8Gb                                                                                                & 3200                                             & x8                                                    & -                                                                                               & -                                                                                                & 10.3K                                                                                         & 1.14e-02                                             & 1.7                                                                                                                                                             & 10.5K                                                                                         & 9.82e-03                                             & 1.7                                                                                                                                                 & 10.5K                                   & 9.82e-03                  \\
                                                                                                                                     & B7                                                                                               & M378A1K43DB2-CTD~\cite{datasheetM378A1K43DB2}                     & 8Gb                                                                                                & 2666                                             & x8                                                    & D                                                                                               & 10-21                                                                                            & 7.3K                                                                                          & 1.32e-01                                             & 2.0                                                                                                                                                             & 7.6K                                                                                          & 1.33e-01                                             & 2.0                                                                                                                                                 & 7.6K                                    & 1.33e-01                  \\
                                                                                                                                     & B8                                                                                               & CMK16GX4M2B3200C16~\cite{datasheetCMK16GX4M2B3200C16}                   & 8Gb                                                                                                & 3200                                             & x8                                                    & -                                                                                               & -                                                                                                & 11.6K                                                                                         & 2.88e-02                                             & 1.7                                                                                                                                                             & 10.5K                                                                                         & 2.37e-02                                             & \agyextthree{1.8}                                                                                                                                                 & 11.\agyextthree{7}K                                   & 2.\agyextthree{5}8e-02                  \\
                                                                                                                                     & B9                                                                                               & M471A5244CB0-CRC\cite{datasheetM471A5244CB0}                     & 8Gb                                                                                                & 2133                                             & x8                                                    & C                                                                                               & 19-19                                                                                            & 11.8K                                                                                         & 2.68e-02                                             & 1.7                                                                                                                                                             & 8.8K                                                                                          & 2.39e-02                                             & \agyextthree{1.8}                                                                                                                                                 & 1\agyextthree{2.3}K                                   & 2.\agyextthree{54}e-02                  \\ 
\hline
\multirow{10}{*}{\rotcell{\begin{tabular}[c]{@{}l@{}}Mfr. C (SK Hynix)\end{tabular}}}                                                             & C0                                                                                               & F4-2400C17S-8GNT~\cite{datasheetF42400C17S8GNT}                     & 4Gb                                                                                                & {2400}         & x8                                                    & B                                                                                               & 02-21                                                                                            & 19.3K                                                                                         & 7.29e-03                                             & 1.7                                                                                                                                                             & 23.4K                                                                                         & 6.61e-03                                             & 1.7                                                                                                                                                 & 23.4K                                   & 6.61e-03                  \\
                                                                                                                                     & C1                                                                                               & F4-2400C17S-8GNT~\cite{datasheetF42400C17S8GNT}                     & 4Gb                                                                                                & {2400}         & x8                                                    & B                                                                                               & 02-21                                                                                            & 19.3K                                                                                         & 6.31e-03                                             & 1.7                                                                                                                                                             & 20.6K                                                                                         & 5.90e-03                                             & 1.7                                                                                                                                                 & 20.6K                                   & 5.90e-03                  \\
                                                                                                                                     & C2                                                                                               & KSM32RD8/16HDR~\cite{datasheetKSM32RD8}                       & 8Gb                                                                                                & 3200                                             & x8                                                    & D                                                                                               & 48-20                                                                                            & 9.6K                                                                                          & 2.82e-02                                             & 1.5                                                                                                                                                             & 9.2K                                                                                          & 2.34e-02                                             & 2.\agyextthree{3}                                                                                                                                                 & \agyextthree{10.0}K                                    & 2.8\agyextthree{9}e-02                  \\
                                                                                                                                     & C3                                                                                               & KSM32RD8/16HDR~\cite{datasheetKSM32RD8}                     & 8Gb                                                                                                & 3200                                             & x8                                                    & D                                                                                               & 48-20                                                                                            & 9.3K                                                                                          & 2.57e-02                                             & 1.5                                                                                                                                                             & 8.9K                                                                                          & 2.21e-02                                             & 2.\agyextthree{3}                                                                                                                                                 & 9.\agyextthree{7}K                                    & 2.\agyextthree{66}e-02                  \\
                                                                                                                                     & C4                                                                                               & HMAA4GU6AJR8N-XN~\cite{datasheetHMAA4GU6AJR8N}                     & 16Gb                                                                                               & 3200                                             & x8                                                    & A                                                                                               & 51-20                                                                                            & 11.6K                                                                                         & 3.22e-02                                             & 1.5                                                                                                                                                             & 11.7K                                                                                         & 2.88e-02                                             & 1.5                                                                                                                                                 & 11.7K                                   & 2.88e-02                  \\
                                                                                                                                     & C5                                                                                               & HMAA4GU6AJR8N-XN~\cite{datasheetHMAA4GU6AJR8N}                     & 16Gb                                                                                               & 3200                                             & x8                                                    & A                                                                                               & 51-20                                                                                            & 9.4K                                                                                          & 3.28e-02                                             & 1.5                                                                                                                                                             & 12.7K                                                                                         & 2.85e-02                                             & 1.5                                                                                                                                                 & 12.7K                                   & 2.85e-02                  \\
                                                                                                                                     & C6                                                                                               & CMV4GX4M1A2133C15~\cite{datasheetCMV4GX4M1A2133C15}                     & 4Gb                                                                                                & 2133                                             & x8                                                    & C                                                                                               & -                                                                                                & 14.2K                                                                                         & 3.08e-02                                             & 1.6                                                                                                                                                             & 15.5K                                                                                         & 2.25e-02                                             & 1.6                                                                                                                                                 & 15.5K                                   & 2.25e-02                  \\
                                                                                                                                     & C7                                                                                               & CMV4GX4M1A2133C15~\cite{datasheetCMV4GX4M1A2133C15}                     & 4Gb                                                                                                & 2133                                             & x8                                                    & C                                                                                               & -                                                                                                & 11.7K                                                                                         & 3.24e-02                                             & 1.6                                                                                                                                                             & 13.6K                                                                                         & 2.60e-02                                             & 1.6                                                                                                                                                 & 13.6K                                   & 2.60e-02                  \\
                                                                                                                                     & C8                                                                                               & KSM32RD8/16HDR~\cite{datasheetKSM32RD8}                       & 8Gb                                                                                                & 3200                                             & x8                                                    & D                                                                                               & 48-20                                                                                            & 11.4K                                                                                         & 2.69e-02                                             & 1.6                                                                                                                                                             & 9.5K                                                                                          & 2.57e-02                                             & 2.5                                                                                                                                                 & 11.4K                                    & 2.69e-02                  \\
                                                                                                                                     & C9                                                                                               & F4-2400C17S-8GNT~\cite{datasheetF42400C17S8GNT}                     & 4Gb                                                                                                & {2400}         & x8                                                    & B                                                                                               & 02-21                                                                                            & 12.6K                                                                                         & 2.18e-02                                             & 1.7                                                                                                                                                             & 15.2K                                                                                         & 1.63e-02                                             & 1.7                                                                                                                                                 & 15.2K                                   & 1.63e-02                  \\
\hline
\end{tabular}
}
\label{tab:detailed_dimm_table}
\end{table}

\end{document}